\documentstyle{livrev}

\RequirePackage{epsf}
\RequirePackage{longtable}

\begin{document}

\title{Characteristic Evolution and Matching}

\author{Jeffrey Winicour\\
Max-Planck-Institut f\" ur Gravitationsphysik,
Albert-Einstein-Institut, \\ 14476 Golm, Germany \\
and Department of Physics and Astronomy,\\
University of Pittsburgh, Pittsburgh, PA 15260, USA\\
jeff@einstein.phyast.pitt.edu\\ \\
{\em (Accepted 10 January 2001)}
}
\date{}
\maketitle

{\bf Abstract}

I review the development of numerical evolution codes for general relativity
based upon the characteristic initial value problem. Progress is traced from
the early stage of 1D feasibility studies to current 3D black codes that
simulate binary black holes. A prime application of characteristic evolution is
Cauchy-characteristic matching, which is also reviewed.

\section{Introduction}
\label{intro}

It is my pleasure to review progress in numerical relativity based upon
characteristic evolution. In the spirit of Living Reviews, I invite my
colleagues to continue to send me contributions and comments at
jeff@einstein.phyast.pitt.edu. 

We have entered an era in which Einstein's equations can effectively be
considered solved at the local level. Several groups, as reported here and in
other Living Reviews, have developed 3D codes which are stable and accurate in
some sufficiently local setting. Global solutions are another matter. In
particular, there is no single code in existence today which purports to be
capable of computing the waveform of gravitational radiation emanating from the
inspiral and merger of two black holes, the premier problem in classical
relativity. Just as several coordinate patches are necessary to describe a
space-time with nontrivial topology, the most effective attack on the binary
black hole problem is likely to involve patching together regions of space-time
handled by different codes.

Much work in numerical relativity is based upon the Cauchy ``3 + 1'' formalism
\cite{york}, with the gravitational radiation extracted by perturbative Cauchy
methods which introduce an artificial Schwarzschild background
\cite{ab1,ab2,ab3,all1}. These wave extraction methods have not been tested in
a nonlinear 3D setting. A different approach which is specifically tailored to
study radiation is based upon the characteristic initial value problem. In the
1960's, Bondi \cite{1bondi,bondi} and Penrose \cite{Penrose}  pioneered the use
of null hypersurfaces to describe gravitational waves. This new approach has
flourished in general relativity. It yields the standard description of the
``plus'' and ``cross'' polarization modes of gravitational radiation in terms
of the real and imaginary parts of the Bondi news function at future null
infinity ${\cal I}^+$.

From a computational standpoint, the major drawback of the characteristic
approach arises from the formation of caustics in the light rays generating the
null hypersurfaces.  In the most ambitious scheme proposed at the theoretical
level such caustics would be treated ``head-on'' as part of the dynamical
problem \cite{friedst1}. This is a profoundly attractive idea. Only a few
structural stable caustics can arise in numerical evolution, and their
geometrical properties are well enough understood to model their singular
behavior numerically \cite{friedst2}. However, a computational implementation
of this approach has not yet been achieved. It is a great idea that is perhaps
ahead of its time.

In the typical setting for the characteristic initial value problem, the domain
of dependence of a single nonsingular null hypersurface is empty. In order to
obtain a nontrivial evolution problem, the null hypersurface must either be
completed to a caustic-crossover region where it pinches off, or an additional
boundary must be introduced. So far, the only caustics that have been
successfully evolved numerically in general relativity are pure point caustics
(the complete null cone problem). When spherical symmetry is not present, it
turns out that the stability conditions near the vertex of a light cone place a
strong restriction on the allowed time step \cite{igw}. Point caustics in
general relativity have been successfully handled this way for axisymmetric
space-times \cite{papa}, but the computational demands for 3D evolution would
be prohibitive using current generation supercomputers. This is unfortunate
because, away from the caustics, the characteristic evolution offers myriad
computational and geometrical advantages.

\begin{figure}[hptb]
  \def\epsfsize#1#2{.5#1}
  \centerline{\epsfbox{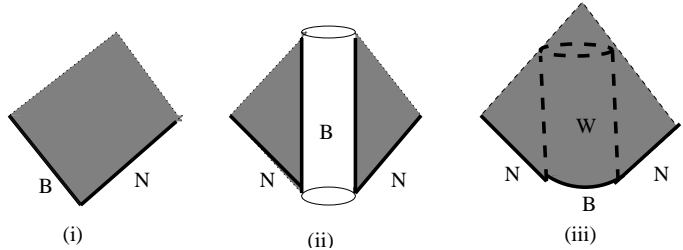}}
  \caption{\it The three applications of characteristic evolution with data
  given on an initial null hypersurface N and boundary B. The shaded regions
  indicate the corresponding domains of dependence.}
  \label{fig:civp}
\end{figure}

As a result, at least in the near future, the computational application of
characteristic evolution is likely to be restricted to some mixed form, in
which data is prescribed on a non-singular but incomplete initial null
hypersurface N and on a second boundary hypersurface B, which together with the
initial null hypersurface determine a nontrivial domain of dependence. This
second hypersurface may itself be either (i) null, (ii) timelike or (iii)
spacelike. These possibilities give rise to the (i) the double null problem,
(ii) the nullcone-worldtube problem or (iii) the Cauchy-characteristic matching
(CCM) problem, in which the Cauchy and characteristic evolutions are matched
transparently across a worldtube W, as schematically depicted in
Fig. \ref{fig:civp}. 

In CCM, it is possible to choose the matching interface between the
Cauchy and characteristic regions to be a null hypersurface, but it is
more practical to match across a timelike worldtube. CCM combines the
advantages of characteristic evolution in treating the outer radiation
zone in spherical coordinates which are naturally adapted to the
topology of the worldtube with the advantages of Cauchy evolution in
Cartesian coordinates in the region where spherical coordinates would
break down.

In this review, we trace the development of characteristic algorithms from
model 1D problems to a 3D code designed to calculate the waveform emitted in
the merger to ringdown phase of a binary black hole. And we trace the
development of CCM from early feasibility studies through current attempts to
treat the binary black hole problem.

There have been several notable developments since my last review. Most
important for future progress has been the award of three new doctorates, with
theses based upon characteristic evolution codes: Denis Pollney~\cite{pollney}
at the the University of Southampton, Luis Lehner~\cite{luisth} and Bela
Szil\'{a}gyi~\cite{belath} at the  University of Pittsburgh. Lehner received
the Nicholas Metropolis award of the American Physical Society for his thesis
research, recognizing characteristic evolution as an important computational
technique. In the area of scientific application, light cone evolution codes
for hydrodynamics have been used to study astrophysical processes. In addition,
a 3D characteristic vacuum code developed at the University of Canberra has
been successfully applied to the scattering of waves off a Schwarzschild black
hole (see Sec.~\ref{sec:3d}) and the Pittsburgh 3D code is being applied to
obtain the postmerger to ringdown waveform from binary black holes (see
Sec.~\ref{sec:nullbbh}). Simulations from these studies can be viewed at the 
Canberra \cite{canberra} and Pittsburgh \cite{pitt} web sites. 

\section{The Characteristic Initial Value Problem}

Characteristics have traditionally played an important role in the
analysis of hyperbolic partial differential equations. However, the use
of characteristic hypersurfaces to supply the foliation underlying an
evolution scheme has been mainly restricted to relativity. This is
perhaps natural because in curved space-time there is no longer a
preferred Cauchy foliation provided by the Euclidean 3-spaces allowed
in Galilean or special relativity. The method of shooting along
characteristics is a standard technique in many areas of computational
physics, but evolution based upon characteristic hypersurfaces is quite
uniquely limited to relativity.

Bondi's initial use of null coordinates to describe radiation
fields \cite{1bondi} was followed by a rapid development of other null
formalisms. These were distinguished either as metric based approaches,
as developed for axisymmetry by Bondi, Metzner and van den
Burg \cite{bondi} and generalized by Sachs \cite{sachs}, or as null
tetrad approaches in which the Bianchi identities appear as part of the
set of equations, as developed by Newman and Penrose \cite{NP}.

At the outset, null formalisms were applied to construct asymptotic
solutions at null infinity by means of $1/r$ expansions. Soon
afterwards, Penrose devised the conformal compactification of null
infinity ${\cal I}$ (``scri''), thereby reducing to geometry the
asymptotic description of the physical properties of the radiation zone,
most notably the Bondi mass and news
function \cite{Penrose}. The characteristic initial value problem
rapidly became an important tool for the clarification of fundamental
conceptual issues regarding gravitational radiation and its energy
content. It laid bare and geometricized the gravitational far field.

The initial focus on asymptotic solutions clarified the kinematic properties of
radiation fields but could not supply the waveform from a specific source. It
was soon realized that instead of carrying out a $1/r$ expansion, one could
reformulate the approach in terms of the integration of ordinary differential
equations along the characteristics (null rays) \cite{tam}. The integration
constants supplied on some inner boundary then determined the specific
waveforms obtained at infinity. In the double-null initial value problem of
Sachs \cite{sachsdn}, the integration constants are supplied at the
intersection of outgoing and ingoing null hypersurfaces. In the
worldtube-nullcone formalism, the sources were represented by integration
constants on the worldtube \cite{tam}. These early formalisms have gone through
much subsequent revamping. Some have been reformulated to fit the changing
styles of modern differential geometry. Some have been reformulated in
preparation for implementation as computational algorithms. See the articles in
\cite{southam} for a representative sample of formalisms.  Rather than
including a review of the extensive literature on characteristic formalisms in
general relativity, I concentrate here on those approaches which have been
implemented as computational evolution schemes. The topic of well-posedness of
the underlying boundary value problems, which has obvious relevance to the
success of numerical simulations, is treated in a separate Living Review by A.
Rendall~\cite{rendall}.

All characteristic evolution schemes share the same skeletal form. The
fundamental ingredient is a foliation by null hypersurfaces $u=const$
which are generated by a 2-dimensional set of null rays, labeled by
coordinates $x^A$, with a coordinate $\lambda$ varying along the rays.
In $(u,\lambda,x^A)$ null coordinates, the main set of Einstein
equations take the schematic form
\begin{equation}
       F_{,\lambda} = H_F[F,G]
\end{equation}
and
\begin{equation}
       G_{,u\lambda} = H_G[F,G,G_{,u}].
\end{equation}
Here $F$ represents a set of hypersurface variables; $G$, a set of evolution
variables; and $H_F$ and $H_G$ are nonlinear hypersurface operators, i.e. they
operate locally on the values of $F$, $G$ and $G_{,u}$ intrinsic to a single
null hypersurface.  In addition to these main equations, there is a subset of
four Einstein equations which are satisfied by virtue of the Bianchi
identities, provided that they are satisfied on a hypersurface transverse to
the characteristics. These equations have the physical interpretation as
conservation laws. Mathematically they are analogous to the constraint
equations of the canonical formalism. But they are not elliptic, since they are
imposed upon null or timelike hypersurfaces, rather than spacelike Cauchy
hypersurfaces.

\section{Characteristic Evolution Codes}

Computational implementation of characteristic evolution may be based upon
different versions of the formalism (i.e. metric or tetrad) and different
versions of the initial value problem (i.e. double null or worldtube-nullcone).
The performance and computational requirements of the resulting evolution codes
can vary drastically. However, most characteristic evolution codes share certain
common advantages:

(i) There are no elliptic constraint equations. This eliminates the
need for time consuming iterative methods to enforce constraints.

(ii) No second time derivatives appear so that the number of basic
variables is at least half the number for the corresponding version of
the Cauchy problem.

(iii) The main Einstein equations form a system of coupled ordinary
differential equations with respect to the parameter $\lambda$ varying
along the characteristics. This allows construction of an evolution
algorithm in terms of a simple march along the characteristics.

(iv) In problems with isolated sources, the radiation zone can be
compactified into a finite grid boundary using Penrose's conformal
technique. Because the Penrose boundary is a null hypersurface, no
extraneous outgoing radiation condition or other artificial boundary
condition is required.

(v) The grid domain is exactly the region in which waves propagate,
which is ideally efficient for radiation studies. Since each null
hypersurface of the foliation extends to infinity, the radiation is
calculated immediately (in retarded time). 

(vi) In black hole space-times, a large redshift at null infinity
relative to internal sources is an indication of the formation of an
event horizon and can be used to limit the evolution to the exterior
region of spacetime.

Characteristic schemes also share as a common disadvantage the necessity either
to deal with caustics or to avoid them altogether. The scheme to tackle the
caustics head on by including their development as part of the evolution is
perhaps a great idea still ahead of its time but one that should not be
forgotten. There are only a handful of structurally stable caustics, and they
have well known algebraic properties. This makes it possible to model their
singular structure in terms of Pad\' e approximants. The structural stability of
the singularities should in principle make this possible, and algorithms to
evolve the elementary caustics have been proposed \cite{cstew,padstew}. In the
axisymmetric case, cusps and folds are the only stable caustics, and they have
already been identified in the horizon formation occurring in simulations of
head-on collisions of black holes and in the temporarily toroidal horizons
occurring in collapse of rotating matter \cite{sci,torus}. In a generic binary
black hole horizon, where axisymmetry is broken, there is a closed curve of
cusps which bounds the 2-dimensional region on the horizon where the black
holes initially form and merge~\cite{ndata,asym}.

\subsection{1+1 Dimensional Codes}
\label{sec:1d}

It is often said that the solution of the general ordinary
differential equation is essentially known, in light of the success of
computational algorithms and present day computing power. Perhaps this
is an overstatement because investigating singular behavior is still an
art. But, in the same vein, it is fair to say that the general system
of hyperbolic partial differential equations in one spatial dimension
is a solved problem. At least, it seems to be true in general
relativity.

One of the earliest characteristic evolution codes, constructed by Corkill and
Stewart \cite{cstew,bonn}, treated space-times with two Killing vectors using a
grid  based upon double null coordinates, with the null hypersurfaces
intersecting in the surfaces spanned by the Killing vectors. They simulated
colliding plane waves and evolved the Khan-Penrose \cite{khan} collision of
impulsive ($\delta$-function curvature) plane waves to within a few numerical
zones from the final singularity, with extremely close agreement with the
analytic results.  Their simulations of collisions with more general waveforms,
for which exact solutions are not known, provided input to the understanding of
singularity formation which was unforseen in the analytic treatments of this
problem.

Many 1+1 dimensional characteristic codes have been developed for spherically
symmetric systems. Here matter must be included in order to make the system
non-Schwarzschild.   Initially the characteristic evolution of matter was
restricted to simple cases, such as massless Klein-Gordon fields, which allowed
simulation of gravitational collapse and radiation effects in the simple
context of spherical symmetry. Now, characteristic evolution of matter is
progressing to more complicated systems. Its application to hydrodynamics,
which is beginning to make significant contributions to general relativistic
astrophysics, is reviewed in Sec.~\ref{sec:grace}.

The synergy between analytic and computational approaches has already led to
dramatic results in the massless Klein-Gordon case. On the analytic side,
working in a characteristic initial value formulation based upon outgoing null
cones, Christodoulou made a penetrating study of the  existence and uniqueness
of solutions to this problem. \cite{X1986,X21986,X1987,X21987} He showed that
weak initial data evolve to Minkowski space asymptotically in time, but that
sufficiently strong data form a horizon, with nonzero Bondi mass. In the latter
case, he showed that the geometry is asymptotically Schwarzschild in the
approach to $I^+$ (future time infinity) from outside the horizon, thus
establishing a rigorous version of the no-hair theorem. What this analytic
tour-de-force did not reveal was the remarkable critical behavior in the
transition between these two regimes, which was discovered by Choptuik
\cite{chsouth,choptprl} by computational simulation based upon Cauchy
evolution. This phenomena has now been understood in terms of the methods of
renormalization group theory and intermediate asymptotics and has spawned a new
subfield in general relativity, which is covered in a Living Review by C.
Gundlach \cite{gundlach}.

The characteristic evolution algorithm for the spherically symmetric
Einstein-Klein-Gordon problem provides a simple illustration of the techniques
used in the general case. It centers about the evolution scheme for the scalar
field, which constitutes the only dynamical field. Given the scalar field, all
gravitational quantities can be determined by integration along the
characteristics of the null foliation. This is a coupled problem, since the
scalar wave equation involves the curved space metric. It illustrates how null
algorithms lead to a hierarchy of equations which can be integrated along the
characteristics to effectively decouple the hypersurface and dynamical
variables. 

In a Bondi coordinate system based upon outgoing null hypersurfaces
$u=const$ and a surface area coordinate $r$, the metric is
\begin{equation}     
    ds^2= -e^{2\beta}du({V \over r}du+2dr) 
    + r^2(d\theta^2 +\sin^2\theta d\phi^2). \label{eq:metric}
\end{equation}     
Smoothness at $r=0$ allows the coordinate conditions 
\begin{equation}
    V(u,r)=r+O(r^3)\quad {\rm and}\quad \beta(u,r)=O(r^2).   
\label{eq:bc}
\end{equation}
The field equations consist of the wave equation $\Box \Phi = 0$ for
the scalar field and two hypersurface equations for the metric
functions:
\begin{equation}
     \beta_{,r}= 2\pi r(\Phi_{,r})^2    \label{eq:sbeta}
\end{equation}
\begin{equation}
     V_{,r}=e^{2\beta}.             \label{eq:sv}
\end{equation}
The wave equation can be expressed in the form
\begin{equation}
     \Box ^{(2)}g-({V\over r})_{,r}{e^{-2\beta} g\over r}=0,
\label{eq:hatwave}
\end{equation} 

where $g=r\Phi$ and $\Box ^{(2)}$ is the D'Alembertian associated with
the two dimensional submanifold spanned by the ingoing and outgoing
null geodesics. Initial null data for evolution consists of
$\Phi(u_0,r)$ at initial retarded time $u_0$.

\begin{figure}[hptb]
  \def\epsfsize#1#2{.5#1} \centerline{\epsfbox{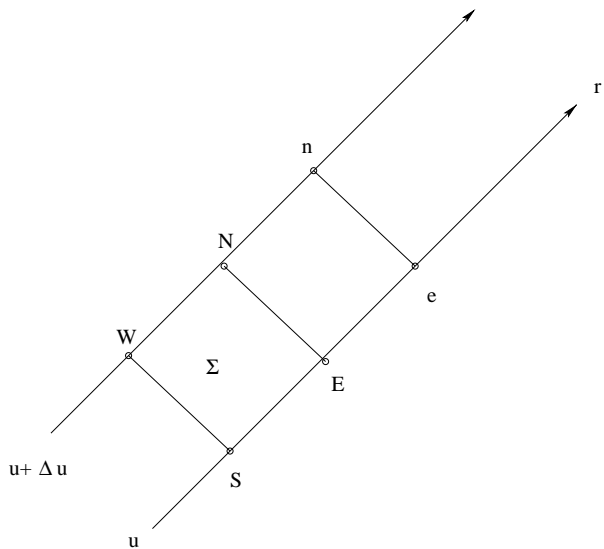}} \caption{\it The null
  parallelogram. After computing the field at point $N$, the algorithm marches
  the computation to ${\cal I}^+$ by shifting the corners by $N\rightarrow n$,
  $E\rightarrow e$, $S\rightarrow E$, $W\rightarrow N$.}
  \label{fig:nsew}
\end{figure}

Because any two dimensional geometry is conformally flat, the surface integral
of $\Box^{(2)} g$ over a null parallelogram $\Sigma$ gives exactly the same
result as in a flat 2-space, and leads to an integral identity upon which a
simple evolution algorithm can be based \cite{EKG}. Let the vertices of the
null parallelogram be labeled by $N$, $E$, $S$ and $W$ corresponding,
respectively, to their relative locations North, East, South and West in the
2-space, as shown in Fig. \ref{fig:nsew}. Upon integration of
(\ref{eq:hatwave}), curvature introduces an area integral correction to the
flat space null parallelogram relation between the values of $g$ at the
vertices:
\begin{equation}
     g_N - g_W - g_E + g_S =
     -{1 \over 2} \int_\Sigma du dr ({V\over r})_{,r} {g\over r} .
\label{eq:integral}
\end{equation}

This identity, in one form or another, lies behind all of the null
evolution algorithms that have been applied to this system. The prime
distinction between the different algorithms is whether they are based
upon double null coordinates or Bondi coordinates as in Eq.
(\ref{eq:metric}). When a double null coordinate system is adopted, the
points  $N$, $E$, $S$ and $W$ can be located in each computational cell
at grid points, so that evaluation of the left hand side of Eq.
(\ref{eq:integral}) requires no interpolation. As a result, in flat
space, where the right hand side of Eq. (\ref{eq:integral}) vanishes,
it is possible to formulate an exact evolution algorithm. In curved
space, of course, there is truncation error arising from the
approximation of the integral by evaluating the integrand at the center
of $\Sigma$.

The identity (\ref{eq:integral}) gives rise to the following explicit marching
algorithm, indicated in Fig. \ref{fig:nsew}. Let the null parallelogram lie at
some fixed $\theta$ and $\phi$ and span adjacent retarded time levels $u_0$ and
$u_0+\Delta u$. Imagine for now that the points $N$, $E$, $S$ and $W$ lie on
the spatial grid, with $r_N-r_W=r_E-r_S=\Delta r$. If $g$ has been determined
on the initial cone $u_0$, which contains the points $E$ and $S$, and radially
outward from the origin to the point $W$ on the next cone $u_0+\Delta u$, then
Eq. (\ref{eq:integral}) determines $g$ at the next radial grid point $N$ in
terms of an integral over $\Sigma$. The integrand can be approximated to second
order, i.e. to $O(\Delta r \Delta u)$, by evaluating it at the center of
$\Sigma$. To this same accuracy, the value of $g$ at the center equals its
average between the points $E$ and $W$, at which $g$ has already been
determined.  Similarly, the value of $(V/r)_{,r}$ at the center of $\Sigma$ can
be approximated to second order in terms of values of $V$ at points where it
can be determined by integrating the hypersurface equations (\ref{eq:sbeta})
and (\ref{eq:sv}) radially outward from $r=0$.

After carrying out this procedure to evaluate $g$ at the point $N$, the
procedure can then be iterated to determine $g$ at the next radially outward
grid point on the $u_0+\Delta u$ level, i.e. point $n$ in Fig. \ref{fig:nsew}.
Upon completing this radial march to null infinity, in terms of a compactified
radial coordinate such as $x=r/(1+r)$, the field $g$ is then evaluated on the
next null cone at $u_0+2\Delta u$, beginning at the vertex where smoothness
gives the startup condition that $g(u,0)=0$.

In the compactified Bondi formalism, the vertices $N$, $E$, $S$ and $W$
of the null parallelogram $\Sigma$ cannot be chosen to lie exactly on
the grid because, even in Minkowski space, the velocity of light in
terms of a compactified radial coordinate $x$ is not constant. As a
consequence, the fields $g$, $\beta$ and $V$ at the vertices of
$\Sigma$ are approximated to second order accuracy by interpolating
between grid points.  However, cancellations arise between these four
interpolations so that Eq.(\ref{eq:integral}) is satisfied to fourth
order accuracy.  The net result is that the finite difference version
of (\ref{eq:integral}) steps $g$ radially outward one zone with an
error of fourth order in grid size, $O((\Delta u)^2 (\Delta x)^2)$.  In
addition, the smoothness conditions (\ref{eq:bc}) can be incorporated
into the startup for the numerical integrations for $V$ and $\beta$ to
insure no loss of accuracy in starting up the march at $r=0$. The
resulting global error in $g$, after evolving a finite retarded time,
is then $O(\Delta u\Delta x)$, after compounding errors from $1/(\Delta
u\Delta x)$ number of zones.

When implemented on a grid based upon the $(u,r)$ coordinates, the stability of
this algorithm is subject to a Courant-Friedrichs-Lewy (CFL) condition
requiring that the physical domain of dependence be contained in the numerical
domain of dependence. In the spherically symmetric case, this condition
requires that the ratio of the time step to radial step be limited by
$(V/r)\Delta u \le 2\Delta r$, where $\Delta r=\Delta[x/(1-x)]$. This condition
can be built into the code using the value $V/r=e^{2H}$, corresponding to the
maximum of $V/r$ at $\cal I^+$. The strongest restriction on the time step then
arises just before the formation of a horizon, where $V/r\rightarrow \infty$ at
$\cal I^+$, This infinite redshift provides a mechanism for locating the true
event horizon ``on the fly'' and restricting the evolution to the exterior
space-time.  Points near $\cal I^+$ must be dropped in order to evolve across
the horizon due to the lack of a nonsingular compactified version of future
time infinity  $I^+$.

The situation is quite different in a double null coordinate system, in which
the vertices of the null parallelogram can be placed exactly on grid points so
that the CFL condition is automatically satisfied. A characteristic code based
upon double null coordinates was used by Goldwirth and Piran in a study of
cosmic censorship \cite{goldw} based upon the spherically symmetric
gravitational collapse of a scalar field. Their early study lacked the
sensitivity of adaptive mesh refinement which later enabled Choptuik to
discover the critical phenomena appearing in this problem. Subsequent work by
Marsa and Choptuik~\cite{mc} combined the use of the null related ingoing
Eddington-Finklestein coordinates with singularity excision to construct a 1D
code that ``runs forever''. Later, Garfinkle~~\cite{garf1} constructed an
improved version of  the Goldwirth-Piran double null code which was able to
simulate critical phenomena without using adaptive mesh refinement. In this
treatment, as the evolution proceeds on one outgoing null cone to the next, the
grid points follow the ingoing null cones and must be dropped as they cross the
origin at $r=0$. However, after half the grid points are lost they are  then
``recycled'' at new positions midway the remaining grid points. This technique
is crucial to resolving the critical phenomena associated with an
$r\rightarrow  0$ sized horizon. An extension of the code~\cite{garf2} was
later used to verify that scalar field collapse in six dimensions continues to
display critical phenomena.

Hamad\'e and Stewart \cite{hamst} also applied a double null code to study
critical phenomena. In order to obtain the accuracy necessary to confirm
Choptuik's results they developed the first example of a characteristic grid
with adaptive mesh refinement (AMR). They did this with both the standard
Berger and Oliger algorithm and their own simplified version, with both
versions giving indistinguishable results. Their simulations of critical
collapse of a scalar field agreed with Choptuik's values for the universal
parameters governing mass scaling and displayed the echoing associated with
discrete self-similarity. Hamad\'e, Horne and Stewart \cite{hamhst} extended
this study to the spherical collapse of an axion/dilaton system and found in
this case that the self-similarity was a continuous symmetry of the critical
solution.

A code based upon Bondi coordinates, developed by S. Husa and his
collaborators~\cite{huscrit}, has been successful applied to spherically
symmetric critical collapse of a nonlinear $\sigma$-model coupled to gravity.
Critical phenomena cannot be resolved on a static grid based upon the Bondi
$r$-coordinate. Instead, the numerical techniques of Garfinkle were adopted by
using a dynamic grid following the ingoing null rays and by recycling radial
grid points. They studied how coupling to gravity affects the critical behavior
previously observed by P. Bizo\' {n}~\cite{bizon} and others in the Minkowski
space version of the model. For a wide range of the coupling constant, they
observe discrete self-similarity and typical mass scaling near the critical
solution.   The code is shown to be second order accurate and to give second
order convergence for the value of the critical parameter.

The first characteristic code in Bondi coordinates for the self gravitating
scalar wave problem was constructed by G\'omez and Winicour \cite{EKG}. They
introduced a numerical compactification of ${\cal I}^+$ for the purpose of
studying  effects of self-gravity on the scalar radiation, particularly in the
high amplitude limit of the rescaling $\Phi\rightarrow a\Phi$. As $a
\rightarrow\infty$, the red shift creates an effective boundary layer at
${\cal I}^+$ which causes the Bondi mass $M_B$ and the scalar field monopole
moment $Q$ to be related by $M_B\sim \pi |Q|/\sqrt{2}$, rather than the
quadratic relation of the weak field limit \cite{scasym}. This could also be
established analytically so that the high amplitude limit provided a check on
the code's ability to handle strongly nonlinear fields. In the small amplitude
case, this work {\it incorrectly} reported that the radiation tails from black
hole formation had an exponential decay characteristic of quasinormal modes
rather than the polynomial $1/t$ or $1/t^2$ falloff expected from Price's
\cite{price} work on perturbations of Schwarzschild black holes. In hindsight,
the error here was not having confidence to run the code sufficiently long to
see the proper late time behavior.

Gundlach, Price and Pullin \cite{gpp1,gpp2} subsequently reexamined the
issue of power law tails using a double null code similar to that developed by
Goldwirth and Piran. Their numerical simulations verified the existence of
power law tails in the full nonlinear case, thus establishing consistency with
analytic perturbative theory. They also found normal mode ringing at
intermediate time, which provided reassuring consistency with perturbation
theory and showed that there is a region of space-time where the results of
linearized theory are remarkably reliable even though highly nonlinear behavior
is taking place elsewhere. These results have led to a methodology that has
application beyond the confines of spherically symmetric problems, most notably
in the ``close approximation'' for the binary black hole problem \cite{pp}.
Power law tails and quasinormal ringing were also confirmed using Cauchy
evolution \cite{mc}.

The study of the radiation tail decay of a scalar field was subsequently
extended by G\'omez, Schmidt and Winicour \cite{gsw} using a characteristic
code. They showed that the Newman-Penrose constant \cite{conserved} for the
scalar field determines the exponent of the power law (not the static monopole
moment as often stated). When this constant is non-zero, the tail decays as
$1/t$, as opposed to the $1/t^2$ decay for the vanishing case.  (There are also
$t^{-n}\log t$ corrections in addition to the exponentially decaying contributions
of the quasinormal modes).  This code was also used to study the instability of
a topological kink in the configuration of the scalar field \cite{kink}. The
kink instability provides the simplest example of the turning point
instability  \cite{ipser,sork} which underlies gravitational collapse of static
equilibria.

The Southampton group has constructed a 1+1 dimensional characteristic code for
spacetimes with cylindrical symmetry \cite{cylinder1,cylinder2}. The original
motivation was to use it as the exterior characteristic code in a test case of
Cauchy-characteristic matching (CCM). (See Sec.~\ref{sec:cylmatch} of this
review for the application to matching.) However, U. Sperhake, K.R.P..
Sj\" odin and J.~A. Vickers~\cite{vick1,vick2} have recently modified
the code into a global characteristic version for the purpose of studying
cosmic strings, represented by massive scalar and vector fields coupled to
gravity. Using a Geroch decomposition~\cite{gerdec} with respect to the
translational Killing vector, the global problem is reduced to a 2+1
dimensional asymptotically flat spacetime, so that ${\cal I}^+$ can be
compactified and included in the numerical grid. Rather than the explicit
scheme used in CCM, the new version employs an implicit, second order in space
and time, Crank-Nicholson evolution scheme. The code showed long term stability
and second order convergence in vacuum tests based upon exact Weber-Wheeler
waves~\cite{wweb}, Xanthopoulos' rotating solution~\cite{xanth} and in tests of
wave scattering by a string. The results show damped ringing of
the string after a incoming Weber-Wheeler pulse has excited it and scattered
out to ${\cal I}^+$. The ringing frequencies are independent of the details of
the pulse but are inversely proportional to the masses of the scalar and
vector fields.

The group at the Universidad de Oriente in Venezuela has applied characteristic
methods to study the self-similar collapse of spherical matter and charge
distributions~\cite{barreto96,barreto98,barreto99}. The assumption of
self-similarity reduces the problem to a system of ODE's, subject to boundary
conditions determined by matching to an exterior Reissner-Nordstrom-Vaidya
solution. Heat flow in the internal fluid is balanced at the surface by the
Vaidya radiation. Their solutions illustrate how a nonzero total charge can halt
gravitational collapse and produce a final stable equilibrium~\cite{barreto99}.
It is interesting that the pressure vanishes in the final equilibrium state so
that hydrostatic support is completely supplied by Coulomb repulsion.

\subsection{2 + 1 Codes}

One dimensional characteristic codes enjoy a very special simplicity
due to the two preferred sets (ingoing and outgoing) of characteristic
null hypersurfaces. This eliminates a source of gauge freedom that
otherwise exists in either two or three dimensional characteristic
codes.  However, the manner in which characteristics of a hyperbolic
system determine domains of dependence and lead to propagation
equations for shock waves is the same as in the one dimensional
case. This makes it desirable for the purpose of numerical evolution to
enforce propagation along characteristics as extensively as possible.
In basing a Cauchy algorithm upon shooting along characteristics, the
infinity of characteristic rays (technically, {\it bicharacteristics})
at each point leads to an arbitrariness which, for a practical
numerical scheme, makes it necessary either to average the propagation
equations over the sphere of characteristic directions or to select out
some preferred subset of propagation equations. The latter
approach was successfully applied by Butler \cite{butler} to the
Cauchy evolution of 2-dimensional fluid flow but there seems to have
been very little follow-up along these lines.

The formal ideas behind the construction of two or three dimensional
characteristic codes are the same, although there are different technical
choices of angular coordinates for the null rays. Historically, most
characteristic work graduated first from 1D to 2D because of the available
computing power.

\subsection{The Bondi Problem}

The first characteristic code based upon the original Bondi equations for a
twist-free axisymmetric space-time was constructed by Isaacson, Welling and
Winicour \cite{isaac} at Pittsburgh. The space-time was foliated by a family of
null cones, complete with point vertices at which regularity conditions were
imposed. The code accurately integrated the hypersurface and evolution
equations out to compactified null infinity. This allowed studies of the Bondi
mass and radiation flux on the initial null cone, but it could not be used as a
practical evolution code because of an instability at the vertices of the null
cones.

These instabilities came as a rude shock and led to a retreat to the
simpler problem of axisymmetric scalar waves propagating in Minkowski
space, with the metric
\begin{equation}
        ds^2= -du^2-2du dr +r^2(d\theta^2 +\sin^2\theta d\phi^2) 
\label{eq:mink}
\end{equation} in outgoing null cone coordinates.  A null cone code for
this problem was constructed using an algorithm based upon
Eq.(\ref{eq:integral}), with the angular part of the flat space
Laplacian replacing the curvature terms in the integrand on the right
hand side. This simple setting allowed the instability to be traced to
a subtle violation of the CFL condition near the vertices of the cones.
In terms of grid spacings $\Delta x^{\alpha}$, the CFL condition in
this coordinate system takes the explicit form
\begin{equation}
   \Delta u/ \Delta r < -1 +[K^2+(K-1)^2-2K(K-1)\cos \Delta \theta]^{1/2}, 
\label{eq:cfl}
\end{equation}
where the coefficient $K$,  which is of order 1, depends on the
particular startup procedure adopted for the outward integration.  Far
from the vertex, the condition (\ref{eq:cfl}) on the time step $\Delta
u$ is quantitatively similar to the CFL condition for a standard Cauchy
evolution algorithm in spherical coordinates. But condition
(\ref{eq:cfl}) is strongest near the vertex of the cone where (at the
equator $\theta =\pi/2$) it implies that
\begin{equation} 
    \Delta u < K \Delta r (\Delta \theta)^2.
\end{equation}
This is in contrast to the analogous requirement
\begin{equation} 
\Delta u < K\Delta r \Delta \theta,
\end{equation}
for stable Cauchy evolution near the origin of a spherical coordinate
system. The extra power of $\Delta \theta$ is the price that must be
paid near the vertex for the simplicity of a characteristic code.
Nevertheless, the enforcement of this condition allowed efficient
global simulation of axisymmetric scalar waves. Global studies of
backscattering, radiative tail decay and solitons were carried out
for nonlinear axisymmetric waves \cite{isaac}, but 3-dimensional
simulations extending to the vertices of the cones was impractical
on existing machines.

Aware now of the subtleties of the CFL condition near the vertices, the
Pittsburgh group returned to the Bondi problem, i.e. to evolve the Bondi
metric \cite{bondi}
\begin{eqnarray}
ds^2 & = & ({V \over r}  e ^{2 \beta} 
     - U^2 r^2 e^{2 \gamma}) du^2 
     + 2 e^{2 \beta} du dr 
     + \  2 U r^2 e^{2 \gamma} du d\theta \nonumber \\
     &   & - r^2 (e^{2 \gamma} d\theta^2
        + e^{-2\gamma} \sin^2 \theta d\phi^2), 
\label{eq:bmetric}
\end{eqnarray}
by means of the three hypersurface equations
\begin{equation}
\beta_{,r} = \frac{1}{2} r\: (\gamma_{,r})^2  \label{eq:beta}
\end{equation}
\begin{equation}
   [r^4 \, e^{2(\gamma-\beta)} U_{,r}]_{,r} =
   2 r^2 [ r^2  (\frac{\beta}{r^{2}})_{,r\theta} 
   - \frac{(\sin^2 \theta \,\gamma)_{,r\theta}}{\sin^{2}\theta} 
   + 2\,\gamma_{,r} \, \gamma_{,\theta}) ]  \label{eq:U}
\end{equation}
\begin{eqnarray}
   V_{,r} & = &-\frac{1}{4} r^4 e^{2(\gamma-\beta)}(U_{,r})^2 
   + \frac{(r^4 \sin\theta U )_{,r\theta}}{2 r^2 \sin\theta}
   \nonumber \\
   & & +e^{2 (\beta - \gamma)}[ 
       1 - \frac{(\sin \theta \beta_{,\theta})_{,\theta}}{\sin\theta} 
       + \gamma _{,\theta\theta} + 3 \cot\theta \gamma_{,\theta} 
       - (\beta_{,\theta})^2  \nonumber \\
     &  &-  2 \gamma_{,\theta} (\gamma_{,\theta} -\beta_{,\theta}) ]            
\label{eq:V}
\end{eqnarray}
and the evolution equation
\begin{eqnarray}
4 r ( r \gamma)_{,ur}  & = &  [ 2 r \gamma_{,r} V  
- r^{2} ( 2 \gamma_{,\theta}\,U  
          + \sin \theta ({U \over {\sin \theta}})_{,\theta} )]_{,r} 
-2 r^{2} \frac{(\gamma_{,r} U \sin\theta)_{,\theta}}{\sin\theta} 
\nonumber \\
& & + \frac{1}{2} r^{4} e^{2(\gamma-\beta)} (U_{,r})^{2} 
+ 2 e^{2(\beta - \gamma)} [ (\beta_{,\theta})^2 
     + \sin \theta ({\beta_{,\theta} \over {\sin \theta}})_{,\theta} ] .  
\label{eq:gammaev}
\end{eqnarray}

The beauty of the Bondi equations is that they form a clean hierarchy. Given
$\gamma$ on an initial null hypersurface, the equations can be integrated
radially to determine $\beta$, $U$, $V$ and $\gamma_{,u}$ on the hypersurface
(in that order) in terms of integration constants determined by boundary
conditions, or smoothness if extended to the vertex of a null cone.  The
initial data $\gamma$ is unconstrained except by smoothness conditions. Because
$\gamma$ represents a spin-2 field, it must be $O(\sin^2 \theta)$ near the
poles of the spherical coordinates and must consist of $l\ge 2$ spin-2
multipoles. 

In the computational implementation of this system by the Pittsburgh
group~\cite{papa}, the null hypersurfaces were chosen to be complete null cones
with nonsingular vertices, which (for simplicity) trace out a geodesic
worldline $r=0$. The smoothness conditions at the vertices were formulated in
local Minkowski coordinates. 

The vertices of the cones were not the chief source of
difficulty. A null parallelogram marching algorithm, similar to that
used in the scalar case, gave rise to an instability that sprang up
throughout the grid. In order to reveal the source of the instability,
physical considerations suggested looking at the linearized version of
the Bondi equations, since they must be related to the wave equation.
If this relationship were sufficiently simple, then the scalar wave
algorithm could be used as a guide in stabilizing the evolution of
$\gamma$. A scheme for relating $\gamma$ to solutions $\Phi$ of the
wave equation had been formulated in the original paper by Bondi,
Metzner and van den Burgh \cite{bondi}.  However, in that scheme, the
relationship of the scalar wave to $\gamma$ was nonlocal in the
angular directions and was not useful for the stability analysis.

A local relationship between $\gamma$ and solutions of the wave
equation was found \cite{papa}. This provided a test bed for the null
evolution algorithm similar to the Cauchy test bed provided by
Teukolsky waves \cite{teuk}. More critically, it allowed a simple von
Neumann linear stability analysis of the finite difference equations,
which revealed that the evolution would be unstable if the metric
quantity $U$ was evaluated on the grid. For a stable
algorithm, the grid points for $U$ must be staggered between the grid
points for $\gamma$, $\beta$ and $V$. This unexpected feature
emphasizes the value of linear stability analysis in formulating stable
finite difference approximations.

It led to an axisymmetric code for the global Bondi
problem which ran stably, subject to a CFL condition, throughout the
regime in which caustics and horizons did not form. Stability in this
regime was verified experimentally by running arbitrary initial data
until it radiated away to ${\cal I}^+$. Also, new exact solutions as
well as the linearized  null solutions were used to perform extensive
convergence tests that established second order accuracy. The code
generated a large complement of highly accurate numerical solutions for
the class of asymptotically flat, axisymmetric vacuum space-times, a
class for which no analytic solutions are known. All results of
numerical evolutions in this regime were consistent with the theorem of
Christodoulou and Klainerman \cite{XKlain} that weak initial data
evolve asymptotically to Minkowski space at late time.

An additional global check on accuracy was performed  using Bondi's
formula relating mass loss to the time integral of the square of the
news function. The Bondi mass loss formula is not one of the equations
used in the evolution algorithm but follows from those equations as a
consequence of a global integration of the Bianchi identities.  Thus it
not only furnishes a valuable tool for physical interpretation but it
also provides a very important calibration of numerical accuracy and
consistency.

An interesting feature of the evolution arises in regard to
compactification.  By construction, the $u$-direction is timelike at
the origin where it coincides with the worldline traced out by the
vertices of the outgoing null cones. But even for weak fields, the
$u$-direction generically becomes spacelike at large distances along an
outgoing ray. Geometrically, this reflects the property that ${\cal I}$
is itself a null hypersurface so that all internal directions are
spacelike, except for the null generator. For a flat space time, the
$u$-direction picked out at the origin leads to a null evolution
direction at ${\cal I}$, but this direction becomes spacelike under a
slight deviation from spherical symmetry. Thus the evolution
generically becomes ``superluminal'' near ${\cal I}$.  Remarkably,
there were no adverse numerical effects. This fortuitous property
apparently arises from the natural way that causality is built into the
marching algorithm so that no additional resort to numerical
techniques, such as ``causal differencing'' \cite{Alliance97b}, was
necessary.

\subsubsection{The Conformal-Null Tetrad Approach}

J. Stewart has implemented a characteristic evolution code which handles the
Bondi problem by a null tetrad, as opposed to metric, formalism \cite{stewbm}.
The geometrical algorithm underlying the evolution scheme, as outlined in
\cite{friedst1,friedst2}, is  H. Friedrich's \cite{fried} conformal-null
description of a compactified space-time in terms of a first order system of
partial differential equations. The variables include the metric, the
connection, and the curvature, as in a Newman-Penrose formalism, but in
addition the conformal factor (necessary for compactification of ${\cal I}$)
and its gradient. Without assuming any symmetry, there are more than 7 times as
many variables as in a metric based null scheme, and the corresponding
equations do not decompose into as clean a hierarchy. This disadvantage,
compared to the metric approach, is balanced by several advantages: (1) The
equations form a symmetric hyperbolic system so that standard theorems can be
used to establish that the system is well-posed. (2) Standard evolution
algorithms can be invoked to ensure numerical stability. (3) The extra
variables associated with the curvature tensor are not completely excess
baggage, since they supply essential physical information. (4) The
regularization necessary to treat ${\cal I}$ is built in as part of the
formalism so that no special numerical regularization techniques are necessary
as in the metric case. (This last advantage is somewhat offset by the necessity
of having to locate ${\cal I}$ by tracking the zeroes of the conformal factor.)

The code was intended to study gravitational waves from an axisymmetric star.
Since only the vacuum equations are evolved, the outgoing radiation from the
star is represented by data ($\Psi_4$ in Newman-Penrose notation) on an ingoing
null cone forming the inner boundary of the evolved domain. The inner boundary
data is supplemented by Schwarzschild data on the initial outgoing null cone,
which models an initially quiescent state of the star. This provides the
necessary data for a double-null initial value problem. The evolution would
normally break down where the ingoing null hypersurface develops caustics. But
by choosing a scenario in which a black hole is formed, it is possible to
evolve the entire region exterior to the horizon. An obvious test bed is the
Schwarzschild space-time for which a numerically satisfactory evolution was
achieved (convergence tests were not reported).

Physically interesting results were obtained by choosing data corresponding to
an outgoing quadrupole pulse of radiation. By increasing the initial amplitude
of the data $\Psi_4$, it was possible to evolve into a regime where the energy
loss due to radiation was large enough to drive the total Bondi mass negative.
Although such data is too grossly exaggerated to be consistent with an
astrophysically realistic source, the formation of a negative mass is an
impressive test of the robustness of the code.

\subsubsection{Twisting Axisymmetry}
\label{sec:axiev}

The Southampton group, as part of its long range goal of combining Cauchy and
characteristic evolution has developed a code~\cite{south1,south2,pollney}
which extends the Bondi problem to full axisymmetry, as described by the
general characteristic formalism of Sachs~\cite{sachs}. By dropping the
requirement that the rotational Killing vector be twist-free, they are able to
include rotational effects, including radiation in the ``cross'' polarization
mode (only the ``plus'' mode is allowed by twist-free axisymmetry). The null
equations and variables are recast into a suitably regularized form to allow
compactification of null infinity. Regularization at the vertices or caustics
of the null hypersurfaces is not necessary, since they anticipate matching to
an interior Cauchy evolution across a finite worldtube. 

The code is designed to insure Bondi coordinate conditions at infinity, so that
the metric has the asymptotically Minkowskian form corresponding to
null-spherical coordinates. In order to achieve this, the hypersurface equation
for the Bondi metric variable $\beta$ must be integrated radially inward from
infinity, where the integration constant is specified. The evolution of the
dynamical variables proceeds radially outward as dictated by
causality~\cite{pollney}. This differs from the Pittsburgh code in which all
the equations are integrated radially outward, so that the coordinate
conditions are determined at the inner boundary and the metric is
asymptotically flat but not asymptotically Minkowskian. The Southampton scheme
simplifies the formulae for the Bondi news function and mass in terms of the
metric. It is anticipated that the inward integration of $\beta$ causes no
numerical problems because this is a gauge choice which does not propagate
physical information. However, the code has not yet been subject to
convergence and long term stability tests so that these issues cannot be
properly assessed at the present time.

The matching of the Southampton axisymmetric code
to a Cauchy interior is discussed in Sec.~\ref{sec:aximatch}.

\subsection{The Bondi Mass}

Numerical calculations of asymptotic quantities such as the Bondi mass must
overcome severe technical difficulties arising from the necessity to pick off
nonleading terms in an asymptotic expansion about infinity. For example, in an
asymptotically inertial frame (called a Bondi frame at ${\cal I}^+$), the mass
aspect ${\cal M}(u,\theta,\phi)$ must be picked off from the asymptotic
expansion of Bondi's metric quantity $V$ (see Eq. (\ref{eq:V})) of the form $V
= r- 2{\cal M} +O(1/r)$.  This is similar to the experimental task of
determining the mass of an object by measuring its far field. The job is more
difficult if the gauge choice does not correspond to a Bondi frame at ${\cal
I}^+$. One must then deal with an arbitrary foliation of ${\cal I}^+$ into
retarded time slices which are determined by the details of the interior
geometry. As a result, $V$ has the more complicated asymptotic behavior, given
in the axisymmetric case by 
\begin{eqnarray}
 V-r &=& r^2 (L \sin \theta)_{,\theta}/\sin \theta+  r e^{2 (H-K)}
           \Big[  (1-  e^{-2  (H-K)}) \nonumber   \\ 
     &+&  2 (H_{,\theta} \sin \theta)_{,\theta}/\sin \theta 
               +K_{,\theta \theta} + 3 K_{,\theta} \cot
              \theta + 4  (H_{,\theta})^2 \nonumber \\
      & -&  4  H_{,\theta}K_{,\theta} -2  (K_{,\theta})^2 \Big] - 2 e^{2 H}
               {\cal M} + O(r^{-1}) , 
\label{eq:wasym} 
\end{eqnarray}  
where $L$, $H$ and $K$ are gauge dependent functions of $(u,\theta)$
which would vanish in a Bondi frame \cite{tam,isaac}. The calculation
of the Bondi mass requires regularization of this expression by
numerical techniques so that the coefficient ${\cal M}$ can be picked
off. The task is now similar to the experimental determination of the
mass of an object by using non-inertial instruments in a far zone which
contains $O(1/r)$ radiation fields. But it can be done!

It was accomplished in Stewart's code by reexpressing the formula for
the Bondi mass in terms of the well-behaved fields of the conformal
formalism \cite{stewbm}. In the Pittsburgh code, it was accomplished by
re-expressing the Bondi mass in terms of renormalized metric variables
which regularize all calculations at ${\cal I}^+$ and make them second
order accurate in grid size~\cite{mbondi}. The calculation of the Bondi news
function (which provides the waveforms of both polarization modes) is
an easier numerical task than the Bondi mass. It has also been
implemented in both of these codes, thus allowing the important check
of the Bondi mass loss formula.

An alternative approach to computing the Bondi mass is to adopt a gauge which
corresponds more closely to an inertial or Bondi frame at ${\cal I}^+$ and
simplifies the asymptotic limit. Such a choice is the null quasi-spherical
gauge in which the angular part of the metric is proportional to the unit
sphere metric and as a result the gauge term $K$ vanishes in
Eq.~(\ref{eq:wasym}). This gauge was adopted by the Canberra group in
developing a 3-D characteristic evolution code~\cite{bartnumeth} (see Sec.
\ref{sec:3d} for further discussion). It allowed accurate  computation of the
Bondi mass as a limit as $r \rightarrow\infty$ of the Hawking
mass~\cite{bartint}.

Mainstream astrophysics is couched in Newtonian concepts, some of which
have no well defined extension to general relativity.  In order to
provide a sound basis for relativistic astrophysics, it is crucial to
develop general relativistic concepts which have well defined and
useful Newtonian limits. Mass and radiation flux are
fundamental in this regard. The results of characteristic codes show
that the energy of a radiating system can be evaluated rigorously and
accurately according to the rules for asymptotically flat space-times,
while avoiding the deficiencies that plagued the ``pre-numerical'' era
of relativity:  (i) the use of coordinate dependent concepts such as
gravitational energy-momentum pseudotensors; (ii) a rather loose notion
of asymptotic flatness, particularly for radiative space-times; (iii)
the appearance of divergent integrals; and (iv) the use of
approximation formalisms, such as weak field or slow motion
expansions, whose errors have not been rigorously estimated.

Characteristic codes have extended the role of the Bondi mass from that of a
geometrical construct in the theory of isolated systems to that
of a highly accurate computational tool. The Bondi mass loss formula
provides an important global check on the preservation of the Bianchi
identities. The mass loss rates themselves have important astrophysical
significance. The numerical results demonstrate that computational
approaches, rigorously based upon the geometrical definition of mass in
general relativity, can be used to calculate radiation losses in highly
nonlinear processes where perturbation calculations would not be
meaningful.

Numerical calculation of the Bondi mass has been used to explore both
the Newtonian and the strong field limits of general
relativity \cite{mbondi}. For a quasi-Newtonian system of radiating
dust, the numerical calculation joins smoothly on to a post-Newtonian
expansion of the energy in powers of $1/c$, beginning with the
Newtonian mass and mechanical energy as the leading terms. This
comparison with perturbation theory has been carried out to $O(1/c^7)$,
at which stage the computed Bondi mass peels away from the
post-Newtonian expansion. It remains strictly positive, in contrast to
the truncated post-Newtonian behavior which leads to negative values.

A subtle feature of the Bondi mass stems from its role as one component of the
total energy-momentum 4-vector, whose calculation requires identification of
the translation subgroup of the Bondi-Metzner-Sachs group \cite{bms}. This
introduces boost freedom into the problem. Identifying the translation subgroup
is tantamount to knowing the conformal transformation to a conformal Bondi
frame \cite{tam} in which the time slices of ${\cal I}$ have unit sphere
geometry. Both Stewart's code and the Pittsburgh code adapt the coordinates to
simplify the description of the interior sources. This results in an arbitrary
foliation of ${\cal I}$.  The determination of the conformal factor which
relates the 2-metric $h_{AB}$ of a slice of ${\cal I}$ to the unit sphere
metric is an elliptic problem equivalent to solving the second order partial
differential equation for the conformal transformation of Gaussian curvature.
In the axisymmetric case, the PDE reduces to an ODE with respect to the angle
$\theta$, and is straightforward to solve \cite{mbondi}. The integration
constants determine the boost freedom along the axis of symmetry.

The non-axisymmetric case is more complicated. Stewart~\cite{stewbm}
proposes an approach based upon the dyad decomposition
\begin{equation}
        h_{AB}dx^A dx^B =  m_A dx^A {\bar m}_B  dx^B.  
\end{equation} 
The desired conformal transformation is obtained by first relating
$h_{AB}$ conformally to the flat metric of the complex plane. Denoting
the complex coordinate of the plane by $\zeta$, this relationship can
be expressed as $d\zeta = e^f m_A dx^A$. The conformal factor $f$ can
then be determined from the integrability condition \begin{equation}
           m_{[A} \partial_{B]} f =\partial_{[A}m_{B]}.
\end{equation}
This is equivalent to the classic Beltrami equation for finding
isothermal coordinates. It would appear to be a more effective scheme
than tackling the second order PDE directly, but numerical
implementation has not yet been carried out.

\subsection{3D Characteristic Evolution}
\label{sec:3d}

There has been rapid progress in 3D characteristic evolution. There are now two
independent 3D codes, one developed at Canberra and the other at Pittsburgh,
(the PITT code) which have the capability to study gravitational waves in
single black hole spacetimes, at a level still not mastered by Cauchy codes.
Three years ago the Pittsburgh group established robust stability and second
order accuracy of a fully nonlinear 3D code which calculates waveforms at null
infinity \cite{cce,high} and also tracks a dynamical black hole and excise its
internal singularity from the computational grid~\cite{excise,wobb}. The
Canberra group has implemented an independent nonlinear 3D code which
accurately evolves the exterior region of a Schwarzschild black hole. Both
codes pose data on a an initial null hypersurface and on a worldtube boundary
and evolve the exterior spacetime out to a compactified version of null
infinity, where the waveform is computed.  However, there are essential
differences in the underlying geometrical formalisms and numerical techniques
used in the two codes and their success in evolving generic black hole
spacetimes.

\subsubsection{Geometrical formalism}

The PITT code uses a standard Bondi-Sachs null coordinate system whereas the
Canberra code employs a null quasi-spherical (NQS) gauge (not to be confused
with the quasi-spherical approximation in which quadratically aspherical terms
are ignored~\cite{cce}). The NQS gauge takes advantage of the possibility of
mapping the angular part of the Bondi metric conformally onto a unit sphere
metric, so that $h_{AB}\rightarrow q_{AB}$. The required transformation $x^A
\rightarrow y^A(u,r,x^A)$ is in general dependent upon $u$ and $r$ so that the
quasi-spherical angular coordinates $y^A$ are not constant along the outgoing
null rays, unlike the Bondi-Sachs angular coordinates. Instead the coordinates
$y^A$ display the analogue of a shift on the null hypersurfaces $u=const$. The
radiation content of the metric is contained in a shear vector describing this
shift. This results in the description of the radiation in terms of a
spin-weight 1 field, rather than the spin-weight 2 field associated with
$h_{AB}$ in the Bondi-Sachs formalism. In both the Bondi-Sachs and
NQS gauges, the independent gravitational data on a null
hypersurface is the conformal part of its degenerate 3-metric. The Bondi-Sachs
null data consists of $h_{AB}$, which determines the intrinsic conformal metric
of the null hypersurface. In the quasi-spherical case, $h_{AB}=q_{AB}$ and the
shear vector comprises the only non-trivial part of the conformal 3-metric.
Both the Bondi-Sachs and NQS gauges can be arranged to coincide in the special
case of shear-free Robinson-Trautman metrics~\cite{derry,bartgauge}.

The formulation of Einstein's equations in the NQS gauge is
presented in Ref.~\cite{bartee} and the associated gauge freedom arising from
$(u,r)$ dependent rotation and boosts of the unit sphere is discussed in
Ref.~\cite{bartgauge}. As in the PITT code, the main equations involve
integrating a hierarchy of hypersurface equations along the radial null
geodesics extending from the inner boundary to null infinity. In the
NQS gauge the source terms for these radial ODE's are rather
simple when the unknowns are chosen to be the connection coefficients.
However, as a price to pay for this simplicity, after the radial integrations
are performed on each null hypersurface a first order elliptic equation must
be solved on each $r=const$ cross-section to reconstruct the underlying metric.

\subsubsection{Numerical Methodology}

The PITT code is an explicit second order finite difference evolution algorithm
based upon retarded time steps on a uniform 3-dimensional null coordinate grid.
The straightforward numerical approach and the second order convergence of the
finite difference equations has facilitated code development. The Canberra code
uses an assortment of novel and elegant numerical methods. Most of these
involve smoothing or filtering and have obvious advantage for removing short
wavelength noise but would be unsuitable for modeling shocks.

Both codes require the ability to handle tensor fields and their derivatives on
the sphere. Spherical coordinates and spherical harmonics are natural analytic
tools for the description of radiation but their implementation in
computational work requires dealing with the impossibility of smoothly covering
the sphere with a single coordinate grid.  Polar coordinate singularities in
axisymmetric systems can be regularized by standard tricks. In the absence of
symmetry, these techniques do not generalize and would be especially
prohibitive to develop for tensor fields. 

A crucial ingredient of the PITT code is the {\it eth}-module~\cite{competh}
which incorporates a computational version of the Newman-Penrose
eth-formalism~\cite{eth}. The eth-module covers the sphere with two overlapping
stereographic coordinate grids (North and South). It provides everywhere
regular, second order accurate, finite difference expressions for tensor fields
on the sphere and their covariant derivatives. The eth-calculus
simplifies the underlying equations, avoids spurious coordinate singularities 
and allows accurate differentiation of tensor fields on the sphere in a
computationally efficient and clean way. Its main weakness is the numerical
noise introduced by interpolations (4th order accurate) between the North and
South patches. For parabolic or elliptic equations on the sphere, the finite
difference approach of the eth calculus would be less efficient than a spectral
approach, but no parabolic or elliptic equations appear in the Bondi-Sachs
evolution scheme.

The Canberra code handles fields on the sphere by means of a 3-fold representation~\cite{bartsphere}: (i) discretized functions on a
spherical grid uniformly spaced in standard $(\theta,\phi)$ coordinates, (ii)
fast-Fourier transforms with respect to $(\theta,\phi)$ (based upon a smooth
map of the torus onto the sphere) and (iii) spectral decomposition of  scalar,
vector and tensor fields in terms of spin-weighted spherical harmonics.  The
grid values are used in carrying out nonlinear algebraic operations; the
Fourier representation is used to calculate $(\theta,\phi)$-derivatives; and
the spherical harmonic representation is used to solve global problems, such as
the solution of the first order elliptic equation for the reconstruction of the
metric, whose unique solution requires pinning down the $\ell=1$ gauge freedom.
The sizes of the grid and of the Fourier and spherical harmonic representations
are coordinated. In practice, the spherical harmonic expansion is  carried out
to 15th order in $\ell$ but the resulting coefficients must then be projected
into the $\ell \le 10$ subspace in order to avoid inconsistencies between the
spherical harmonic and the grid and Fourier representations.

The Canberra code solves the null hypersurface equations by combining an 8th
order Runge-Kutta integration with a convolution spline to interpolate field
values. The radial grid points are dynamically positioned to approximate
ingoing null geodesics, a technique originally due to Goldwirth and
Piran~\cite{goldw} to avoid inaccuracies near a horizon resulting from a
uniform $r$-grid due to the coordinate singularity in the case of a stationary
horizon.  The time evolution uses the method of lines with a 4th order
Runge-Kutta integrator, which introduces further high frequency filtering.

\subsubsection{Stability}

{\bf PITT Code}: Analytic stability analysis  of the finite difference
equations has been crucial in the development of a stable evolution algorithm,
subject to the standard Courant-Friedrichs-Levy (CFL) condition for an explicit 
code. Linear stability analysis on Minkowski and Schwarzschild
backgrounds showed that certain field variables must be represented on the
half-grid~\cite{papa,cce}. Nonlinear stability analysis was essential in
revealing and curing a mode coupling instability that was not present in the 
original axisymmetric version of the code~\cite{high,luisdis}. Stability
persists even in the regime that the $u$-direction, along which the grid flows,
becomes spacelike, such as outside the velocity of light cone in a rotating
coordinate system. Severe tests were used to verify stability. In the linear
regime, {\bf robust stability} was established by imposing random initial data
on the initial characteristic hypersurface and random constraint violating
boundary data on an inner worldtube. The code ran stably for 10,000 grid
crossing times under these conditions~\cite{cce,cauchboun}. Except for a
linearized version of a Arnowitt-Deser-Misner code~\cite{cauchboun} (see also
Sec. \ref{sec:ccm}), the PITT code is the only 3D general relativistic code
which has passed this test of robust stability. The use of random data is only
possible in sufficiently weak cases where effective energy terms quadratic in
the field gradients are not dominant. Stability in the highly nonlinear regime
was tested in two ways. Runs for a time of $60,000M$ were carried out for a
moving, distorted Schwarzschild black hole (of mass M), with the marginally
trapped surface at the inner boundary tracked and its interior excised from the
computational grid~\cite{wobb,stablett}. This is the longest black hole
simulation carried out to date. Furthermore, the scattering of a gravitational
wave off a Schwarzschild black hole was successful carried out in the extreme
nonlinear regime where the the backscattered Bondi news was  as large as
$N=400$ (in dimensionless geometric units)~\cite{high}, showing that the code
can cope with the enormous power output $N^2 c^5/G \approx 10^{60}W$ in
conventional units.  This exceeds the power that would be produced if, in 1
second, the entire galaxy were converted to gravitational radiation.

{\bf Canberra code}: Direct stability analysis of the underlying finite
difference equations is impractical because of the extensive mix of spectral
techniques, higher order methods and splines. Similarly, there is no clear-cut
CFL limit on the code, although stability tests show that there is a limit on
the time step. The damping of high frequency modes due to the implicit
filtering would be expected to suppress numerical instability but the stability
of the Canberra code is nevertheless subject to two
qualifications~\cite{bartacc,bartnumsol,bartnumeth}: (i) At late times (less
than 100M), the evolution terminates as it approaches an event horizon,
apparently because of a breakdown of the NQS gauge condition, although an
analysis of how and why this should occur has not yet been given. (ii)
Numerical instabilities arise from dynamic inner boundary conditions and
restrict the inner boundary to a fixed  Schwarzschild horizon. Tests in the
extreme nonlinear regime were not reported.

\subsubsection{Accuracy}

{\bf PITT Code}: Second order accuracy was verified in an extensive number of
testbeds~\cite{cce,high,wobb}, including new exact solutions specifically
constructed in null coordinates for the purpose of convergence tests: 

\begin{itemize}

\item{Linearized waves on a Minkowski background in null cone coordinates}

\item{Boost and rotation symmetric solutions~\cite{boostrot}}

\item{Schwarzschild in rotating coordinates}

\item{Polarization symmetry of nonlinear twist-free axisymmetric waveforms}

\item{Robinson-Trautman waveforms from perturbed Schwarzschild black holes.}

\item{Nonlinear Robinson-Trautman waveforms utilizing an independently
computed solution of the Robinson-Trautman equation.}

\end{itemize}

{\bf Canberra code}: The complexity of the algorithm and NQS gauge makes it
problematic to establish accuracy by direct means. Exact solutions do not
provide an effective convergence check because the Schwarzschild solution is
trivial in the NQS gauge and other known solutions in this gauge require
dynamic inner boundary conditions which destabilize the present version of the
code. Convergence to linearized solutions is a possible check but has not yet
been performed.  Instead indirect tests by means of geometric consistency and
partial convergence tests are used to calibrate accuracy. The consistency
tests are based on the constraint equations, which are not enforced during null
evolution except at the inner boundary. The balance between mass loss and
radiation flux through ${\cal I}^+$ is a global version of these constraints.
No appreciable growth of the constraints is noticeable until within $5M$ of the
final breakdown of the code. In weak field tests where angular resolution does
not dominate the error, partial convergence tests based upon varying the radial
grid size show the 8th order convergence in the shear expected from the
Runge-Kutta integration and splines. When the radial source of error is
small, reduced error with smaller time step can also be discerned.

In practical runs, the major source of inaccuracy is the spherical harmonic
resolution, currently restricted to $\ell \le 15$ by hardware limitations.
Truncation of the spherical harmonic expansion has the effect of modifying the
equations to a system for which the constraints are no longer satisfied. The
relative error in the constraints is $10^{-3}$ for strong field
simulations~\cite{bartint}.

\subsubsection{Nonlinear Scattering Off a Schwarzschild Black Hole}

A natural physical application of a characteristic evolution code is the
nonlinear version of the classic problem of scattering off a Schwarzschild
black hole, first solved perturbatively by Price \cite{price}. Here the inner
worldtube for the characteristic initial value problem consists of the ingoing
$r=2m$ surface (the past horizon), where Schwarzschild data is prescribed. The
nonlinear problem of a gravitational wave scattering off a Schwarzschild black
hole is then posed in terms of data on an outgoing null cone consisting of an
incoming pulse with compact support. Part of the energy of this pulse falls
into the black hole and part is backscattered out to ${\cal I}^+$. This problem
has been investigated using both the PITT and Canberra codes.

The Pittsburgh group studied the backscattered waveform (described by the Bondi
news function) as a function of incoming pulse amplitude. The computational eth
module smoothly handled the complicated time dependent transformation between
the non-inertial computational frame at ${\cal I}^+$ and the inertial (Bondi)
frame necessary to obtain the standard ``plus'' and ``cross'' polarization
modes. In the perturbative regime, the news corresponds to the backscattering
of the incoming pulse off the effective Schwarzschild potential. When the
energy of the pulse is no larger than the central Schwarzschild mass,  the
backscattered waveform  still depends roughly linearly linearly on the
amplitude of the incoming pulse.  However, for very high amplitudes the
waveform behaves quite differently. Its amplitude is greater than that
predicted by linear scaling and its shape drastically changes and exhibits
extra oscillations. In this very high amplitude case, the mass of the system is
completely dominated by the incoming pulse, which essentially backscatters off
itself in a nonlinear way.

The Canberra code was used to study the change in Bondi mass due to
radiation~\cite{bartint}. The Hawking mass $M_H(u,r)$ is first calculated as a
function of radius and retarded time, with the Bondi mass $M_B(u)$ then
obtained by taking the limit $r\rightarrow \infty$. The limit has good
numerical behavior. For a strong initial pulse with $\ell=4$ angular
dependence, in a run from $u=0$ to $u=70$ (in units where the interior
Schwarzschild mass is 1), the Bondi mass drops from 1.8 to 1.00002, showing
that almost half of the initial energy of the system was backscattered and that
a surprisingly negligible amount falls into the black hole. The Bondi mass
decreases monotonically in time, as theoretically necessary, but its rate of
change exhibits an interesting pulsing behavior whose time scale cannot be
obviously explained in terms of quasinormal oscillations. The Bondi mass loss
formula is confirmed with relative error of less than $10^{-3} $. This is
impressive accuracy considering the potential sources of numerical error
introduced by taking the limit of the Hawking mass. The Canberra group also
studied the effect of initial pulse amplitude on the waveform of the
backscattered radiation but did not extend their study to the very high
amplitude regime in which qualitatively interesting nonlinear effects occur.

\subsubsection{Black Hole in a Box}

The PITT code has also be implemented to evolve along a family of {\em ingoing}
null cones, with data given on a worldtube at their {\em outer} boundary and on
an initial {\em ingoing} null cone. The code was used to evolve a black hole in
the region interior to the worldtube by locating the marginally trapped surface
(MTS) on the ingoing cones and excising its singular interior~\cite{excise}.
The code tracks the motion of the MTS and measures its area during the evolution. It was used to simulate a distorted ``black hole in a
box''~\cite{wobb}. Data at the outer worldtube was induced from a Schwarzschild
or Kerr spacetime but the worldtube was allowed to move relative to the
stationary trajectories; i.e. with respect to the grid the worldtube is fixed
but the black hole moves inside it. The initial null data consisted of a pulse
of radiation which subsequently travels outward to the worldtube where it
reflects back toward the black hole. The approach of the system to equilibrium
was monitored by the area of the MTS, which also equals its Hawking mass.
When the worldtube is stationary (static or rotating in place), the distorted
black hole inside evolves to equilibrium with the boundary. A boost or other
motion of the worldtube with respect to the black hole does not affect this
result.  The marginally trapped surface always reaches equilibrium with the
outer boundary, confirming that the motion of the boundary is ``pure gauge''.

The code essentially runs ``forever'' even when the worldtube wobbles with
respect to the black hole to produce artificial periodic time dependence. An
initially distorted, wobbling black hole was evolved for a time of $60,000M$,
longer by orders of magnitude than stability permits for any other existing 3D
black hole code. This exceptional performance opens a promising new approach to
handle the inner boundary condition for Cauchy evolution of black holes by the
matching methods reviewed in Sec \ref{sec:ccm}.

\subsection{Characteristic Treatment of Binary Black Holes}
\label{sec:nullbbh}

An important application of characteristic evolution is the calculation of the
waveform emitted by binary black holes, which is possible during the very
interesting nonlinear domain from merger to ringdown~\cite{ndata,kyoto}. The
evolution is carried out along a family of ingoing null hypersurfaces which
intersect the horizon in topological spheres. It is restricted to the period
following the merger, for otherwise the ingoing null hypersurfaces would
intersect the horizon in disjoint pieces corresponding to the individual black
holes. The evolution proceeds {\it backward} in time on an ingoing null
foliation to determine the exterior spacetime in the post-merger era.  It is an
example of the characteristic initial value problem posed on an intersecting
pair of null hypersurfaces~\cite{sachsdn,haywdn}, for which existence theorems
apply in some neighborhood of the initial null
hypersurfaces~\cite{hagenseifert77,helmut81a,fried}. Here one of the null
hypersurfaces is the event horizon ${\cal H}$ of the binary black holes. The
other is an ingoing null hypersurface $J^+$ which intersects ${\cal H}$ in a
topologically spherical surface ${\cal S}^+$ approximating the equilibrium of
the final Kerr black hole, so that $J^+$ approximates future null infinity
${\cal I}^+$. The required data for the analytic problem consists of the
degenerate conformal null metrics of ${\cal H}$ and $J^+$ and the metric and
extrinsic curvature of ${\cal S}^+$.

The conformal metric of ${\cal H}$ is provided by the conformal horizon model
for a binary black hole horizon~\cite{ndata,asym}, which treats the horizon in
stand-alone fashion as a 3-dimensional manifold endowed with a degenerate
metric $\gamma_{ab}$ and affine parameter $t$ along its null rays. The metric
is obtained from the conformal mapping $\gamma_{ab}=\Omega^2 \hat \gamma_{ab}$
of the intrinsic metric $\hat \gamma_{ab}$ of a flat space null hypersurface
emanating from a convex surface ${\cal S}_0$ embedded at constant time in
Minkowski space.  The horizon is identified with the null hypersurface formed
by the inner branch of the boundary of the past of ${\cal S}_0$, and its
extension into the future. The flat space null hypersurface expands forever as
its affine parameter $\hat t$ (given by Minkowski time) increases but the
conformal factor is chosen to stop the expansion so that the cross-sectional
area of the black hole approaches a finite limit in the future. At the same
time, the Raychaudhuri equation (which governs the growth of surface area)
forces a nonlinear relation between the affine parameters $t$ and $\hat t$
which produces the nontrivial topology of the affine slices of the black hole
horizon. The relative distortion between the affine parameters $t$ and $\hat
t$, brought about by curved space focusing, gives rise to the trousers shape of
a binary black hole horizon. 

\begin{figure}[hptb]
  \def\epsfsize#1#2{.4#1}
  \centerline{\epsfbox{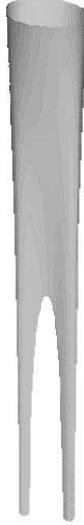}}
  \caption{\it Trousers shaped event horizon obtained by the
  conformal model.}
  \label{fig:pants}
\end{figure}

\newpage

\begin{figure}[hptb]
  \def\epsfsize#1#2{.2#1}
  \centerline{\epsfbox{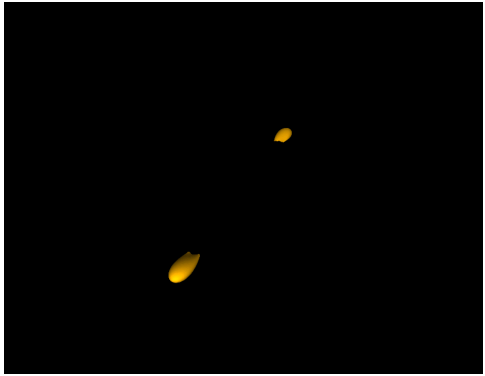}}
  \caption{\it Tidal distortion of approaching black holes.}
  \label{fig:bbh2-30}
\end{figure}

\begin{figure}[hptb]
  \def\epsfsize#1#2{.2#1}
  \centerline{\epsfbox{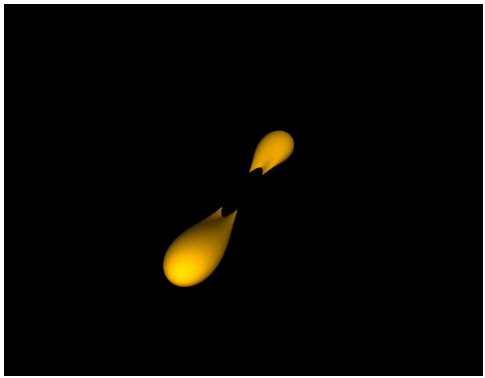}}
  \caption{\it Formation of sharp pincers just prior to merger.}
  \label{fig:bbh2-11}
\end{figure}

\begin{figure}[hptb]
  \def\epsfsize#1#2{.2#1}
  \centerline{\epsfbox{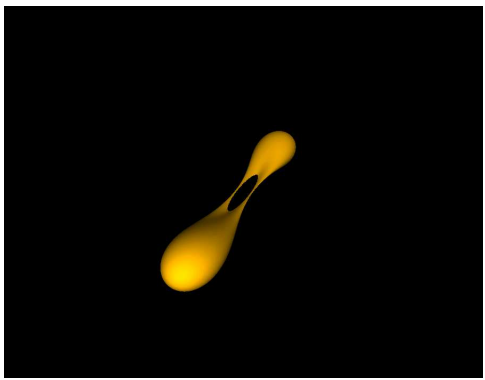}}
  \caption{\it Temporarily toridal stage just after merger.}
  \label{fig:bbh2-10}
\end{figure}

\begin{figure}[hptb]
  \def\epsfsize#1#2{.2#1}
  \centerline{\epsfbox{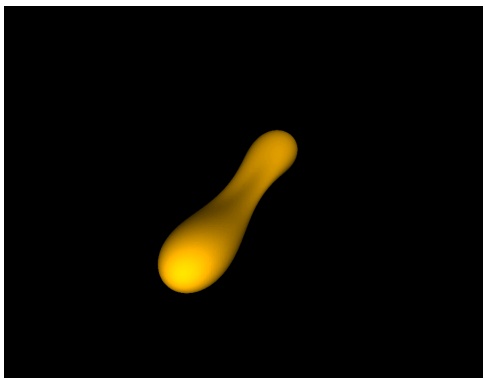}}
  \caption{\it Peanut shape black hole after the hole in the torus closes.}
  \label{fig:bbh2-08}
\end{figure}

\begin{figure}[hptb]
  \def\epsfsize#1#2{.2#1}
  \centerline{\epsfbox{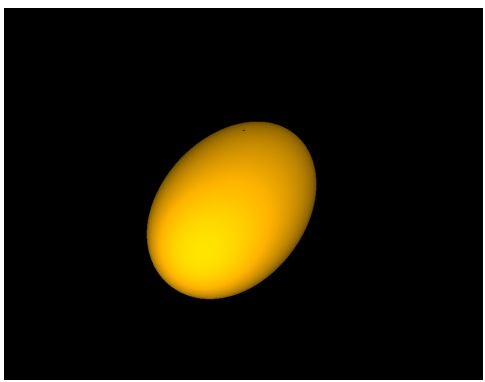}}
  \caption{\it Approach to final equilibrium.}
  \label{fig:bbh2-02}
\end{figure}

An embedding diagram of the horizon for an axisymmetric head-on collision,
obtained by choosing ${\cal S}_0$ to be a prolate spheroid, is shown in Fig.
\ref{fig:pants} \cite{ndata}. The black hole event horizon associated with a
triaxial ellipsoid  reveals new features not seen in the  degenerate case of
the head-on collision~\cite{asym}, as depicted in Fig's. \ref{fig:bbh2-30} -
\ref{fig:bbh2-02}. If the degeneracy is slightly broken, the individual black
holes form with spherical topology but as they approach, tidal distortion
produces two sharp pincers on each black hole just prior to merger. At merger,
the two pincers join to form a single temporarily toroidal black hole. The
inner hole of the torus subsequently closes up (superluminally) to produce
first a peanut shaped black hole and finally a spherical black hole. In the
degenerate axisymmetric limit, the pincers reduce to a point so that the
individual holes have teardrop shape and they merge without a toroidal phase.
No violation of the topological censorship~\cite{fsw} occurs because the hole
in the torus closes up superluminally. Consequently, a causal curve passing
through the torus at a given time can be slipped below the bottom of a trouser
leg to yield a causal curve lying entirely outside the hole~\cite{toroid}.
Details of this merger can be viewed at \cite{pitt}. 

\newpage

The conformal horizon model determines the data on ${\cal H}$ and ${\cal S}^+$.
The remaining data necessary to evolve the exterior spacetime is the conformal
geometry of $J^+$, which constitutes the outgoing radiation waveform. The
determination of the merger-ringdown waveform proceeds in two stages. In the
first stage, this outgoing waveform is set to zero and the spacetime is evolved
backward in time to calculate the incoming radiation entering from ${\cal
I}^-$. (This incoming radiation is eventually absorbed by the black hole.) From
a time reversed point of view, this evolution describes the outgoing waveform
emitted in the fission of a white hole, with the physically correct initial
condition of no ingoing radiation. Preliminary calculations  show that at late
times the waveform is entirely quadrupole ($\ell =2$) but that a strong $\ell
=4$ mode exists just before fission. In the second stage of the calculation,
which has not yet been carried out, this waveform is used to generate the
physically correct outgoing waveform for a black hole merger. The passage from
the first stage to the second is the nonlinear equivalent of first determining
an inhomogeneous solution to a linear problem and then adding the appropriate
homogeneous solution to satisfy the boundary conditions. In this context, the
first stage supplies an advanced solution and the second stage the homogeneous
retarded minus advanced solution. When the evolution is carried out in the
perturbative regime of a Kerr or Schwarzschild background, as in the close
approximation~\cite{jorge}, this superposition of solutions is simplified by
the time reflection symmetry~\cite{kyoto}. More generally, beyond the
perturbative regime, the merger-ringdown waveform must be obtained by a more
complicated inverse scattering procedure. 

There is a complication in applying the PITT code to this double null evolution
because a dynamic horizon does not lie precisely on $r$-grid points. As a
result, the $r$-derivative of the null data, i.e. the ingoing shear of ${\cal
H}$, must also be provided in order to initiate the radial hypersurface
integrations. The ingoing shear is part of the free data specified at ${\cal
S}^+$. Its value on ${\cal H}$ can be determined by integrating (backward in
time) a sequence of propagation equations involving the horizon's twist and
ingoing divergence. A horizon code which carries out these integrations has
been tested to give accurate data even beyond the merger~\cite{hdata}.

The code has revealed new global properties of the head-on collision by 
studying a sequence of data for a family of colliding black holes  which
approaches a single Schwarzschild black hole. The resulting perturbed
Schwarzschild horizon provides global insight into the close limit~\cite{pp},
in which the individual black holes holes have joined in the infinite past. A
marginally anti-trapped surface divides the horizon into interior and exterior
regions, analogous to the division of the Schwarzschild horizon by the $r=2M$
bifurcation sphere. In passing from the perturbative to the strongly nonlinear
regime there is a rapid transition in which the individual black holes move
into the exterior portion of the horizon. The data paves the way for the PITT
code to calculate whether this dramatic time dependence of the horizon
produces an equally dramatic waveform.

\section{Cauchy-Characteristic Matching}
\label{sec:ccm}

Characteristic evolution has many advantages over Cauchy evolution. Its one
disadvantage is the existence of either caustics where neighboring
characteristics focus or, a milder version consisting of crossover between two
distinct characteristics. The vertex of a light cone cone is a highly symmetric
caustic which already strongly limits the time step for characteristic
evolution because of the CFL condition. It does not seem possible for a single
characteristic coordinate system to cover the entire exterior region of a
binary black hole spacetime without developing more complicated caustics or
crossovers. This limits the waveform determined by a purely characteristic
evolution to the post merger period.

Cauchy-characteristic matching (CCM) is a way to avoid such limitations by
combining the strong points of characteristic and Cauchy evolution into
a global evolution~\cite{Bis2}. One of the prime goals of computational relativity is the simulation of the inspiral and merger of
binary black holes. Given the appropriate worldtube data for a binary system in its interior, characteristic evolution can supply the exterior spacetime and the radiated waveform. But determination of the worldtube data for a binary requires an interior Cauchy evolution. CCM is designed to solve such global problems.

The potential advantages of CCM over traditional boundary conditions are: (1)
Accurate waveform and polarization properties at infinity; (2) Computational
efficiency for radiation problems in terms of both the grid domain and the
computational algorithm; (3) Elimination of an artificial outer boundary
condition on the Cauchy problem, which eliminates contamination from back
reflection and clarifies the global initial value problem; and (4) A global
picture of the spacetime exterior to the horizon. 

These advantages have been realized in model tests but CCM has not yet been
successful in either axisymmetric or fully 3-dimensional general relativity.
This difficulty may possibly arise from a pathology in the way boundary
conditions have traditionally been applied in the Arnowitt-Deser-Misner
(ADM)~\cite{adm} formulation of the Einstein equations which, at present, is
the only formulation for which CCM has been attempted. 

Instabilities or
inaccuracies introduced at boundaries have emerged as a major
problem common to all ADM code development and have led to pessimism that such codes might be inherently unstable because
of the lack of manifest hyperbolicity in the underlying equations. In
order to shed light on this issue, B.Szil\'{a}gyi~\cite{belath,cauchboun}, as part of his thesis research, carried out a study of ADM
evolution-boundary algorithms in the simple environment of
linearized gravity, where nonlinear sources of physical or numerical
instability are absent and computing time is reduced by a factor of
five by use of a linearized code.
The two main results, for prescribed values of lapse and shift, were:

\begin{itemize}
\item  On analytic grounds, ADM boundary algorithms which supply values
for all components of the metric (or extrinsic curvature) are
inconsistent.

\item Using a consistent boundary algorithm, which only allows free
specification of the transverse-traceless components of the metric with respect
to the boundary, an unconstrained, linearized ADM evolution can be carried out
in a bounded domain for thousands of crossing times with robust stability.
\end{itemize}

The criteria for robust stability is that the initial Cauchy data and free
boundary data be prescribed as random numbers. It is the most severe test of
stability yet carried out in the Cauchy evolution of general relativity.
Similar robust stability tests were previously successfully carried out for
the PITT characteristic code.

CCM cannot work unless the Cauchy code, as well as the characteristic code,
has a robustly stable boundary. This is necessarily so because interpolations
continually introduce short wavelength noise into the neighborhood of the
boundary. Robustness of the Cauchy boundary is a necessary (although not a
sufficient) condition for the successful implementation of CCM. The robustly
stable ADM evolution-boundary algorithm differs from previous approaches and
offers fresh hope for the success of CCM in general relativity.

\subsection{Computational Boundaries}

Boundary conditions are both the most important and the most difficult
part of a theoretical treatment of most physical systems.  Usually,
that's where all the physics is. And, in computational approaches,
that's usually where all the agony is. Computational boundaries for
hyperbolic systems pose special difficulties. Even with an analytic
form of the correct physical boundary condition in hand, there are
seemingly infinitely more unstable numerical implementations than
stable ones.  In general, a well posed problem places more boundary
requirements on the finite difference equations than on the
corresponding partial differential equations. Furthermore, the methods
of linear stability analysis are often more unwieldy to apply to the
boundary than to the interior evolution algorithm.

The von Neumann stability analysis of the interior algorithm linearizes
the equations, while assuming a uniform infinite grid, and checks that
the discrete Fourier modes do not grow exponentially. There is an
additional stability condition that a boundary introduces into this
analysis.  Consider the one dimensional case.  Normally the mode
$e^{kx}$, with $k$ real, is not included in the von Neumann analysis.
However, if there is a boundary to the right on the $x$-axis, one can
legitimately prescribe such a mode (with $k>0$) as initial data, so that
its stability must be checked. In the case of an additional boundary to
the left, the Ryaben'kii-Godunov theory allows separate investigation
of the interior stability and the stability of each individual
boundary \cite{sod}.

The correct physical formulation of any asymptotically flat, radiative
Cauchy problem also requires boundary conditions at infinity. These
conditions must ensure not only that the total energy and the energy loss by
radiation are both finite, but must also ensure the proper $1/r$ asymptotic
falloff of the radiation fields.  However, when treating radiative
systems computationally, an outer boundary must be established
artificially at some large but finite distance in the wave zone, i.e.
many wavelengths from the source. Imposing an accurate radiation
boundary condition at a finite distance is a difficult task even in the
case of a simple radiative system evolving on a fixed geometric
background.  The problem is exacerbated when dealing with Einstein's
equation.

Nowhere is the boundary problem more acute than in the computation of
gravitational radiation produced by black holes. The numerical study of a black
hole space-time by means of a pure Cauchy evolution involves inner as well as
outer grid boundaries.  The inner boundary is necessary to avoid the
topological complications and singularities introduced by a black hole. For
multiple black holes, the inner boundary consists of disjoint pieces. W. Unruh
(see \cite{thornburg1987}) initially suggested the commonly accepted strategy
for Cauchy evolution of black holes. An inner boundary located at (or near) an
apparent horizon is used to excise the singular interior region. Later (see 
Sec.~\ref{sec:bbhib}), I discuss a variation of this strategy based upon
matching to a characteristic evolution in the inner region.

First, consider the outer boundary problem, in which
Cauchy-characteristic matching has a natural application. In the Cauchy
treatment of such a system, the outer grid boundary is located at some
finite distance, normally many wavelengths from the source. Attempts to
use compactified Cauchy hypersurfaces which extend to spatial infinity
have failed because the phase of short wavelength radiation varies
rapidly in spatial directions \cite{Orsz}.  Characteristic evolution
avoids this problem by approaching infinity along phase fronts.

When the system is nonlinear and not amenable to an exact solution, a finite
outer boundary condition must necessarily introduce spurious physical effects
into a Cauchy evolution. The domain of dependence of the initial Cauchy data in
the region spanned by the computational grid would shrink in time along ingoing
characteristics unless data on a worldtube traced out by the outer grid
boundary is included as part of the problem. In order to maintain a causally
sensible evolution, this worldtube data must correctly substitute for the
missing Cauchy data which would have been supplied if the Cauchy hypersurface
had extended to infinity. In a scattering problem, this missing exterior Cauchy
data might, for instance, correspond to an incoming pulse initially outside the
outer boundary. In a problem where the initial radiation fields are confined to
a compact region inside the boundary, this missing Cauchy data is easy to state
when dealing with a constraint free field, such as a scalar field $\Phi$ where
the Cauchy data outside the boundary would be $\Phi_{,t}=0$. However, the
determination of Cauchy data for general relativity is a global elliptic
constraint problem so that there is no well defined scheme to confine it to a
compact region. Furthermore, even if the data were known on a complete initial
hypersurface extending to infinity, it would be a formidable nonlinear problem
to correctly pose the associated data on the outer boundary. The only
formulation of Einstein's equations whose Cauchy initial-boundary value problem
has been shown to have a unique solution is a symmetric hyperbolic formulation
due to Friedrich and Nagy~\cite{friednag}. However, their mathematical
treatment does not suggest how to construct a consistent evolution algorithm.
  
It is common practice in computational physics to impose some artificial
boundary condition (ABC), such as an outgoing radiation condition, in an
attempt to approximate the proper data for the exterior region. This ABC may
cause partial reflection of an outgoing wave back into the system
\cite{Lind,Orsz,Hig86,Ren}, which contaminates the accuracy of the interior
evolution and the calculation of the radiated waveform. Furthermore, nonlinear
waves intrinsically backscatter, which makes it incorrect to try to entirely
eliminate incoming radiation from the outer region. The errors introduced by
these problems are of an analytic origin, essentially independent of
computational discretization. In general, a systematic reduction of this error
can only be achieved by simultaneously refining the discretization and moving
the computational boundary to larger and larger radii. This is computationally
very expensive, especially for three-dimensional simulations.

A traditional outer boundary condition for the wave equation is the
Sommerfeld condition. For a 3D scalar field this takes the form
$g_{,t}+g_{,r}=0$, where $g=r\Phi$.  This condition is {\em exact} for
a linear wave with spherically symmetric data and boundary. In that
case, the exact solution is $g=f_1(t-r)+f_2(t+r)$ and the Sommerfeld
condition eliminates the incoming wave $f_2$.

Much work has been done on formulating boundary conditions, both exact
and approximate, for linear problems in situations that are not
spherically symmetric and in which the Sommerfeld condition would be
inaccurate. These boundary conditions are given various names in the
literature, e.g.  absorbing or non-reflecting.  A variety of successful
implementations of ABC's have been reported for linear problems. See
the recent articles \cite{giv,Ren,tsy,ryab,jcp97} for a general
discussion of ABC's.

Local ABC's have been extensively applied to linear problems with
varying success \cite{Lind,Eng77,Bay80,Tre86,Hig86,Bla88,Jia90}. Some
of these conditions are local approximations to exact integral
representations of the solution in the exterior of the computational
domain \cite{Eng77}, while others are based on approximating the
dispersion relation of the so-called one-way wave equations
\cite{Lind,Tre86}. Higdon \cite{Hig86} showed that this last approach
is essentially equivalent to specifying a finite number of angles of
incidence for which the ABC's yield perfect transmission.  Local ABC's
have also been derived for the linear wave equation by considering the
asymptotic behavior of outgoing solutions \cite{Bay80}, which
generalizes the Sommerfeld outgoing radiation condition. Although such
ABC's are relatively simple to implement and have a low computational
cost, their final accuracy is often limited because the assumptions
made about the behavior of the waves are rarely met in
practice \cite{giv,tsy}.

The disadvantages of local ABC's have led some workers to consider
exact nonlocal boundary conditions based on integral representations of
the infinite domain problem \cite{Tin86,giv,tsy}. Even for problems
where the Green's function is known and easily computed, such
approaches were initially dismissed as impractical \cite{Eng77};
however, the rapid increase in computer power has made it possible to
implement exact nonlocal ABC's for the linear wave equation and
Maxwell's equations in 3D~\cite{deM,kell}.  If properly implemented,
this kind of method can yield numerical solutions which converge to the
exact infinite domain problem in the continuum limit, keeping the
artificial boundary at a fixed distance.  However, due to nonlocality,
the computational cost per time step usually grows at a higher power with
grid size ($O(N^4)$ per time step in three dimensions) than in a local
approach~\cite{giv,deM,tsy}.

The extension of ABC's to {\it nonlinear} problems is much more difficult.
The problem is normally treated by linearizing the region between the
outer boundary and infinity, using either local or nonlocal linear
ABC's \cite{tsy,ryab}. The neglect of the nonlinear terms in this
region introduces an unavoidable error at the analytic level. But even
larger errors are typically introduced in prescribing the outer
boundary data. This is a subtle global problem because the correct
boundary data must correspond to the continuity of fields and their
normal derivatives when extended across the boundary into the
linearized exterior. This is a clear requirement for any consistent
boundary algorithm, since discontinuities in the field or its
derivatives would otherwise act as spurious sheet source on the
boundary, thereby contaminating both the interior and the exterior
evolutions. But the fields and their normal derivatives constitute an
overdetermined set of data for the linearized exterior problem. So it
is necessary to solve a global linearized problem, not just an exterior
one, in order to find the proper data. The designation ``exact ABC'' is
given to an ABC for a nonlinear system whose only error is due to
linearization of the exterior. An exact ABC requires the use of global
techniques, such as the difference potential method, to eliminate back
reflection at the boundary \cite{tsy}.

To date there have been only a few applications of ABC's to strongly
nonlinear problems \cite{giv}. Thompson \cite{Tho87} generalized a
previous nonlinear ABC of Hedstrom \cite{Hed79} to treat 1D and 2D
problems in gas dynamics. These boundary conditions performed poorly in
some situations because of their difficulty in adequately modeling the
field outside the computational domain \cite{Tho87,giv}. Hagstrom and
Hariharan \cite{Hag88} have overcome these difficulties in 1D gas
dynamics by a clever use of Riemann invariants. They proposed a
heuristic generalization of their local ABC to 3D, but this has not yet
been implemented.

In order to reduce the level of approximation at the analytic level,
an artificial boundary for an nonlinear problem must be placed
sufficiently far from the strong-field region.  This sharply increases
the computational cost in multidimensional simulations \cite{Eng77}.
There seems to be no numerical method which converges (as the
discretization is refined) to the infinite domain exact solution of a
strongly nonlinear wave problem in multidimensions, while keeping the
artificial boundary fixed.

Cauchy-characteristic matching is a strategy that eliminates this
nonlinear source of error. In CCM, Cauchy and characteristic evolution
algorithms are pasted together in the neighborhood of a worldtube to
form a global evolution algorithm. The characteristic algorithm
provides an {\em outer} boundary condition for the interior Cauchy
evolution, while the Cauchy algorithm supplies an {\em inner} boundary
condition for the characteristic evolution. The matching worldtube
provides the geometric framework necessary to relate the two
evolutions. The Cauchy foliation slices the worldtube
into spherical coss-sections. The characteristic evolution is based upon the
outgoing null hypersurfaces emanating from these slices, with the
evolution proceeding from one hypersurface to the next by the outward
radial march described earlier. There is no need to truncate space-time
at a finite distance from the sources, since compactification of the
radial null coordinate makes it possible to cover the infinite space
with a finite computational grid. In this way, the true 
waveform may be directly computed by a finite difference algorithm.
Although characteristic evolution has limitations in regions where
caustics develop, it proves to be both accurate and computationally
efficient in the treatment of exterior regions.

CCM evolves a mixed spacelike-null initial value problem in which Cauchy data
is given in a spacelike region bounded by a spherical boundary ${\cal S}$ and
characteristic data is given on a null hypersurface emanating from ${\cal S}$.
The general idea is not entirely new. An early mathematical investigation
combining space-like and characteristic hypersurfaces appears in the work of
Duff~\cite{Duff}. The three chief ingredients for computational implementation
are: (i) a Cauchy evolution module, (ii) a characteristic evolution module and
(iii) a module for matching the Cauchy and characteristic regions across an
interface. The interface is the timelike worldtube which is traced out by the
flow of ${\cal S}$ along the worldlines of the Cauchy evolution, as determined
by the choice of lapse and shift. Matching provides the exchange of data across
the worldtube to allow evolution without any further boundary conditions, as
would be necessary in either a purely Cauchy or a purely characteristic
evolution.

CCM may be formulated as a purely analytic approach, but its advantages
are paramount in the solution of nonlinear problems where analytic
solutions would be impossible. One of the first applications of CCM was
a hybrid numerical-analytical version, initiated by Anderson and Hobill
for the 1D wave equation \cite{Hobill} (see below). There the
characteristic region was restricted to the far field where it was
handled analytically by a linear approximation.

The full potential of CCM lies in a purely numerical treatment of
nonlinear systems where its error converges to zero in the continuum
limit of infinite grid resolution \cite{Bis,Bis2,Clarke}. For high
accuracy, CCM is also by far the most efficient method. For small
target target error $\varepsilon$, it has been shown that the relative
amount of computation required for CCM ($A_{CCM}$) compared to that
required for a pure Cauchy calculation ($A_{C}$) goes to zero,
$A_{CCM}/A_{C} \rightarrow O$ as $\varepsilon \rightarrow
O$ \cite{cce,vishu}. An important factor here is the use of a
compactified characteristic evolution, so that the whole space-time is
represented on a finite grid. From a numerical point of view this means
that the only error made in a calculation of the radiation
waveform at infinity is the controlled error due to the finite
discretization. Accuracy of a Cauchy algorithm which uses an ABC
requires a large grid domain in order to avoid error from nonlinear
effects in the exterior. The computational demands of matching are
small because the interface problem involves one less dimension than
the evolution problem. Because characteristic evolution algorithms are
more efficient than Cauchy algorithms, the efficiency can be further
enhanced by making the matching radius as small as consistent
with avoiding caustics.

At present, the purely computational version of CCM is exclusively the tool of
general relativists who are used to dealing with novel coordinate systems. A
discussion of its potential appears in \cite{Bis}. Only recently
\cite{Clarke,cylinder1,cylinder2,Ccprl} has its practicability been carefully
explored. Research on this topic has been stimulated by the requirements of the
Binary Black Hole Grand Challenge Alliance, where CCM was one of the strategies
being pursued to provide the boundary conditions and determine the radiation
waveform. But I anticipate that its use will eventually spread throughout
computational physics because of its inherent advantages in dealing with
hyperbolic systems, particularly in 3-dimensional problems where efficiency is
desired. A detailed study of the stability and accuracy of CCM for linear and
non-linear wave equations has been presented in Ref.~\cite{jcp97}, illustrating
its potential for a wide range of problems.

\subsection{Perturbative Cauchy-Characteristic Matching}

In numerous purely analytic applications outside of general relativity,
matching techniques have successfully cured pathologies in perturbative
expansions \cite{aliney}. Matching is a strategy for obtaining a global
solution by patching together solutions obtained using different
coordinate systems for different regions. By adopting each coordinate
system to a length scale appropriate to its domain, a globally
convergent perturbation expansion is sometimes possible in cases where
a single coordinate system would fail. Burke showed that matching could
be used to eliminate some of the divergences arising in perturbative
calculations of gravitational radiation \cite{burke}. Kates and Kegles
further showed that the use of a null coordinate system for the
exterior region is essential in perturbation expansions of curved
space radiation fields \cite{kk}. They investigated the perturbative
description of a scalar field on a Schwarzschild background, in which
case the asymptotic behavior of the Schwarzschild light cones differs
drastically from that of the artificial Minkowski light cones used in the
perturbative expansions based upon the flat space Green function. Use
of the Minkowski light cones leads to {\it nonuniformities} in the
expansion of the radiation fields which are eliminated by the use of
exterior coordinates based upon the true light cones. Kates, Anderson,
Kegles and Madonna extended this work to the fully general relativistic
case and reached the same conclusion \cite{kakm}.

Anderson later applied this approach to the slow motion approximation
of a binary system and obtained a derivation of the radiation reaction
effect on the orbital period which avoided some objections to
earlier approaches~\cite{and87}. The use of the true light cones was
also essential in formulating a mathematical theorem that the Bondi
news function satisfies the Einstein quadrupole formula to leading
order in a Newtonian limit \cite{quad}.  Although questions of
mathematical consistency still remain in the perturbative treatment of
gravitational radiation, it is clear that the use of characteristic
methods pushes these problems to a higher perturbative order.
Characteristic techniques are used in present day codes for the Zerilli
equation~\cite{lous}.

\subsection{Analytic-Numerical Matching for Waves}

One of the earliest computational applications of null coordinates in a
matching scheme was a hybrid numerical-analytical version, by Anderson and
Hobill, for the test problem of 1D scalar waves \cite{Hobill,Hobill2,Hobill3}.
Although the nonlinear near field cannot be treated analytically in a general
relativistic radiation problem, the far field can be handled by a perturbative
expansion. Their strategy was to match an inner numerical solution to an
approximate outer analytic solution.

The initial conditions for the exterior solution were fixed by requiring that
the interior sources be stationary prior to some fixed time, As a result, the
exterior analytic solution is causal in the sense that it is stationary in the
past of some null cone. This effectively introduces a condition that eliminates
extraneous incoming radiation from the system in a physically plausible way and
determines the exterior solution uniquely. An overlap region was introduced
between the numerical interior and the analytic exterior. In the overlap, the
numerical solution was matched to the causal analytic solution, resulting in an
evolution that was everywhere causally meaningful.

This is the physically correct approach to a system which is stationary
prior to a fixed time but is nontrivial to generalize, say, to the
problem of radiation from an orbiting  binary. Anderson and Hobill
first tackled the 1D model problem of an oscillator coupled to a
spherically symmetric, flat space, massless scalar field. The numerical
results were in excellent agreement with the exact analytic solution
(which could be obtained globally for this problem).

They extended the model to include spherical scalar waves propagating in a
spherically symmetric curved background. This introduces backscattering which
obscures the concept of a purely outgoing wave. No exact solution exists to
this problem so that an approximation method was necessary to determine the
exterior analytic solution. The approximation was based upon an expansion
parameter controlling the amount of backscatter. In the 0th approximation, the
scale parameter vanishes and the problem reduces to the flat space case which
can be solved exactly. The flat space Green function was then used to generate
higher order corrections.

A key ingredient of this perturbative scheme is that the wave equation is
solved in retarded null coordinates $(u,r)$ for the curved space metric, so
that the proper causality is built into the Green function at each order of
approximation. The transformation from null coordinates $(u,r)$ to Cauchy
coordinates $(t,r)$ is known analytically for this problem. This allows a
simple match between the null and Cauchy solutions at the boundary of the
Cauchy grid. Their scheme is efficient and leads to consistent results in the
region that the numerical and analytic solutions overlap. It is capable of
handling both strong fields and fast motions.

Later, a global, characteristic, numerical study of the self-gravitating
version of this problem, by G\'{o}mez and Winicour, confirmed that the use of
the true null cones is essential in getting the correct radiated waveform
\cite{finw}. For quasi-periodic radiation, the phase of the waveform is
particular sensitive to the truncation of the outer region at a finite
boundary. Although a perturbative estimate would indicate an $O(M/R)$ error,
this error accumulates over many cycles to produce an error of order $\pi$ in
the phase.

Anderson and Hobill proposed that their method be extended to general
relativity by matching a numerical solution to an analytic $1/r$
expansion in null coordinates. However, the only analytic-numerical
matching schemes that have been implemented in general relativity have
been based upon perturbations of a Schwarzschild background using the
standard Schwarzschild time slicing \cite{ab1,ab2,ab3,all1}. It would
be interesting to compare results with an analytic-numeric matching
scheme based upon the true null cones.  However the original proposal by
Anderson and Hobill has not been carried out.

\subsection{Numerical Matching for 1D Gravitational Systems}

The first numerical implementations of CCM were 1D feasibility studies.
These model problems provided a controlled environment for the
development of CCM, in which either exact or independent numerical
solutions were known. The following studies showed that CCM worked like
a charm in a variety of 1D applications - the matched evolutions were
essentially transparent to the presence of the interface.

\subsubsection{Cylindrical Matching}
\label{sec:cylmatch}

The Southampton group chose cylindrically symmetric systems as their model
problem for developing matching techniques. In preliminary work, they showed
how CCM could be consistently carried out for a scalar wave evolving in
Minkowski space-time but expressed in a nontrivial cylindrical coordinate
system \cite{Clarke}.

They then tackled the gravitational problem. They first set up the machinery
necessary for investigating cylindrically symmetric vacuum space-times
\cite{cylinder1}. Although the problem involves only one spatial dimension,
there are two independent modes of polarization. The Cauchy metric was treated
in the Jordan-Ehlers-Kompaneets canonical form, using coordinates
$(t,r,\phi,z)$ adapted to the $(\phi,z)$ cylindrical symmetry. The advantage
here is that $u=t-r$ is then a null coordinate which can be used for the
characteristic evolution. They successfully recast the equations in a suitably
regularized form for the compactification of ${\cal I}^+$ in terms of the
coordinate $y=\sqrt{1/r}$. The simple analytic relationship between Cauchy
coordinates $(t,r)$ and characteristic coordinates $(u,y)$ facilitated the
translation between Cauchy and characteristic variables on the matching
worldtube, given by $r=const$.

They next implemented the scheme as a numerical
code. The interior Cauchy evolution was carried out
using an unconstrained leap frog scheme. It is notable that they report
no problems with instability, which have arisen in other attempts at
unconstrained leapfrog evolution in general relativity. The
characteristic evolution also used a leap frog scheme for the evolution
between retarded time levels $u$, while numerically integrating the
hypersurface equations outward along the characteristics.

The matching interface was located at points common to both the Cauchy
and characteristic grids. In order to update these points by Cauchy
evolution, it was necessary to obtain field values at the Cauchy
``guard'' points which lie outside the worldtube in the characteristic
region. These values were obtained by interpolation from characteristic
grid points (lying on three levels of null hypersurfaces in order to
ensure second order accuracy). Similarly, the boundary data for
starting up the characteristic integration was obtained by
interpolation from Cauchy grid values inside the worldtube.

The matching code was first tested~\cite{cylinder2} using exact Weber-Wheeler
cylindrical waves~\cite{wweb}, which come in from ${\cal I}^-$, pass through
the symmetry axis and expand out to ${\cal I}^+$. The numerical errors were
oscillatory with low growth rate, and second order convergence was confirmed.
Of special importance, little numerical noise was introduced by the interface.
Comparisons of CCM  were made with Cauchy evolutions using a standard outgoing
radiation boundary condition \cite{piran}. At high amplitudes the standard
condition developed a large error very quickly and was competitive only for
weak waves with a large outer boundary. In contrast, the matching code
performed well even with a small matching radius. Some interesting simulations
were presented in which an outgoing wave in one polarization mode collided with
an incoming wave in the other mode, a problem studied earlier by pure Cauchy
evolution~\cite{stark}. The simulations of the collision were qualitatively
similar in these two studies.

The Weber-Wheeler waves contain only one gravitational degree of freedom. The
code was next tested~\cite{south3} using a exact cylindrically symmetric
solutions, due to Piran, Safier and Katz \cite{katz}, which contain both
degrees of freedom. These solutions are singular at ${\cal I}^+$ so that the
code had to be suitably modified.  Relative errors of the various metric
quantities were in the range $10^{-4}$ to $10^{-2}$. The convergence rate of
the numerical solution starts off as second order but diminishes to first order
after long time evolution. However, more recent modifications, made by U.
Sperhake, H. Sj\" odin and J.~A. Vickers after these tests, have produced an
improved version of the characteristic code (see Sec~\ref{sec:1d}).

\subsubsection{Spherical Matching}

A joint collaboration between groups at Pennsylvania State University and the
University of Pittsburgh applied CCM to the Einstein-Klein-Gordon (EKG)
system with spherical symmetry \cite{ekgmat}. This model problem
allowed simulation of black-hole formation as well as wave
propagation.

The geometrical setup is analogous to the cylindrically symmetric problem.
Initial data were specified on the union of a spacelike hypersurface and a null
hypersurface. The evolution used a 3-level Cauchy scheme in the interior and a
2-level characteristic evolution in the compactified exterior. A constrained
Cauchy evolution was adopted because of its earlier success in accurately
simulating scalar wave collapse \cite{choptprl}. Characteristic evolution was
based upon the null parallelogram algorithm Eq.~(\ref{eq:integral}). The
matching between the Cauchy and characteristic foliations was achieved by
imposing continuity conditions on the metric, extrinsic curvature and scalar
field variables, ensuring smoothness of fields and their derivatives across the
matching interface. The extensive analytical and numerical studies of this
system in recent years aided the development of CCM in this non-trivial
geometrical setting without exact solutions by providing basic knowledge of the
expected physical and geometrical behavior.

The CCM code accurate handled wave propagation and
black hole formation for all values of $M/R$ at the matching radius,
with no symptoms of instability or back reflection. Second order
accuracy was established by checking energy conservation.

\subsubsection{Excising 1D Black Holes}

In further developmental work on the EKG model, the Pittsburgh group
used CCM to formulate a new treatment of the inner Cauchy boundary for
a black hole spacetime~\cite{excise}. In the conventional approach, the
inner boundary of the Cauchy evolution is located at an apparent
horizon, which must lie inside (or on) the event horizon
\cite{wald1984}, so that truncation of the interior space-time at the
apparent horizon cannot causally affect the gravitational waves
radiated to infinity. This is the physical rationale behind the
apparent horizon boundary condition. However, instabilities reported
with the conventional approach motivated an alternative treatment.

\begin{figure}[hptb]
  \def\epsfsize#1#2{.5#1} 
  \centerline{\epsfbox{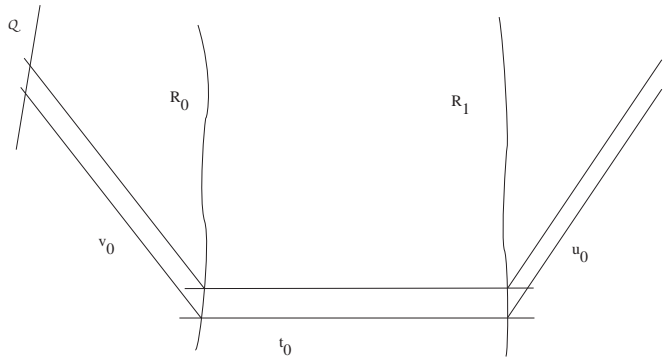}} 
  \caption{\it Black hole excision by matching. A Cauchy evolution, with data
  at $t_0$ is matched across worldtubes $R_0$ and $R_1$ to an ingoing null
  evolution, with data at $v_0$. and an outgoing null evolution, with data at
  $u_0$. The ingoing null evolution extends to an inner trapped boundary $Q$
  and the outgoing null evolution extends to ${\cal I}^+$. } \label{fig:1dexci}
\end{figure}

In the CCM strategy, illustrated in Fig. \ref{fig:1dexci}, the interior black
hole region is evolved using an {\it ingoing} null algorithm whose inner
boundary is a marginally trapped surface and whose outer boundary lies outside
the black hole and forms the inner boundary of a region evolved by the Cauchy
algorithm. In turn, the outer boundary of the Cauchy region is handled by
matching to an outgoing null evolution extending to ${\cal I}^+$. Data is
passed between the inner characteristic and central Cauchy regions using a CCM
procedure similar to that already described for an outer Cauchy boundary. The
main difference is that, whereas the outer Cauchy boundary is matched to an
outgoing null hypersurface, the inner Cauchy boundary is matched to an ingoing
null hypersurface which enters the event horizon and terminates at a marginally
trapped surface.

The translation from an outgoing to an incoming null evolution
algorithm can be easily carried out. The substitution $\beta\rightarrow
\beta+i\pi/2$ in the 3D version of the Bondi metric Eq. (\ref{eq:metric})
provides a simple formal recipe for switching from an outgoing to an
ingoing null formalism~\cite{excise}.

In order to ensure that trapped surfaces exist on the ingoing
null hypersurfaces, initial data was chosen which guarantees black hole
formation. Such data can be obtained from initial Cauchy data for a
black hole. However, rather than extending the Cauchy hypersurface
inward to an apparent horizon, it was truncated sufficiently far outside
the apparent horizon to avoid computational problems with the Cauchy
evolution. The initial Cauchy data was then extended into the black hole
interior as initial null data until a marginally trapped surface is
reached. Two ingredients were essential in order to arrange this.
First, the inner matching surface must be chosen to be convex, in the
sense that its outward null normals uniformly diverge and its inner
null normals uniformly converge. (This is trivial to satisfy in the
spherically symmetric case). Given any physically reasonable matter
source, the focusing theorem then guarantees that the null rays
emanating inward from the matching sphere continue to converge until
reaching a caustic. Second, the initial null data must lead to a
trapped surface before such a caustic is encountered. This is a
relatively easy requirement to satisfy because the initial null data
can be posed freely, without any elliptic or algebraic constraints
other than continuity with the Cauchy data.

A code was developed which implemented CCM at both the inner and outer
boundaries~\cite{excise}. Its performance showed that CCM provides as good a
solution to the black hole excision problem in spherical symmetry as any
previous treatment \cite{scheel1995a,scheel1995b,mc,anninos1995}. CCM is
computationally more efficient than these pure Cauchy approaches (fewer
variables) and much easier to implement. Achieving stability with a pure Cauchy
scheme in the region of an apparent horizon is trickier, involving much trial
and error in choosing finite difference schemes. There were no complications
with stability of the null evolution at the marginally trapped surface.

The Cauchy evolution was carried out in ingoing Eddington-Finklestein
(IEF) coordinates.  The initial Cauchy data consisted of a
Schwarzschild black hole with an ingoing Gaussian pulse of scalar
radiation. Since IEF coordinates are based on ingoing null cones, it
is possible to construct a simple transformation between the IEF Cauchy
metric and the ingoing null metric. Initially there was no scalar field
present on either the ingoing or outgoing null patches. The initial
values for the Bondi variables $\beta$ and $V$ were determined by
matching to the Cauchy data at the matching surfaces and integrating
the hypersurface equations (\ref{eq:sbeta}) and (\ref{eq:sv}).

As the evolution proceeds, the scalar field passes into the black hole, and the
marginally trapped surface (MTS) grows outward. The MTS could be easily located
in the spherically symmetric case by an algebraic equation.  In order to excise
the singular region, the grid points inside the marginally trapped surface were
identified and masked out of the evolution. The backscattered radiation
propagated cleanly across the outer matching surface to ${\cal I}^+$. The
strategy worked smoothly, and second order accuracy of the approach was
established by comparing it to an independent numerical solution obtained using
a second order accurate, purely Cauchy code\cite{mc}. As discussed in
Sec.~\ref{sec:bbhib}, this inside-outside application of CCM has potential
application to the binary black hole problem.

In a variant of this double CCM matching scheme, L. Lehner~\cite{luis2m} has
eliminated the middle Cauchy region and constructed a 1D code matching the
ingoing and outgoing characteristic evolutions directly across a single
timelike worldtube. In this way, he is able to simulate the global problem of a
scalar wave falling into a black hole by purely characteristic methods.

\subsection{Axisymmetric Cauchy-Characteristic Matching}
\label{sec:aximatch}

The Southampton CCM project is being carried out for spacetimes with (twisting)
axial symmetry. The formal basis for the matching scheme was developed by
d'Inverno and Vickers~\cite{south1,south2}.  Similar to the Pittsburgh 3D
strategy (see Sec. \ref{sec:3dccm}, matching is based upon an extraction
module, which supplies boundary data for the exterior characteristic evolution,
and injection module, which supplies boundary data for the interior Cauchy
evolution. However, their use of spherical coordinates for the Cauchy evolution
(as opposed to Cartesian coordinates in the 3D strategy) allows use of a
matching worldtube $r=R_M$ which lies simultaneously on Cauchy and
characteristic gridpoints. This tremendously simplifies the necessary
interpolations between the the Cauchy and characteristic evolutions, at the
expense of dealing with the $r=0$ coordinate singularity in the Cauchy
evolution. The characteristic code, see Sec.~\ref{sec:axiev} is based upon a
compactified Bondi-Sachs formalism. The use of a ``radial'' Cauchy gauge, in
which the Cauchy coordinate $r$ measures the surface area of spheres,
simplifies the relation to the Bondi-Sachs coordinates. In the numerical
scheme, the metric and its derivatives are passed between the Cauchy and
characteristic evolutions exactly at $r=R_M$, thus eliminating the need of a
matching interface encompassing a few grid zones, as in the 3D Pittsburgh
scheme. This avoids a great deal of interpolation error and computational
complexity. 

The implementation of the Southampton CCM algorithm into a running code is
nearing completion, with preliminary results described by D. Pollney in his
thesis~\cite{pollney}. The Cauchy code is based upon the axisymmetric ADM code
of Stark and Piran~\cite{starkpir} and reproduces their vacuum results for
a short time period, after which an instability at the origin becomes manifest.
The characteristic code has been tested to reproduce accurately the
Schwarzschild and boost-rotation symmetric solutions~\cite{boostrot}, with more
thorough tests of stability and accuracy still being carried out.  Much
progress has been made but much work remains to make the code useful for
scientific application.

\subsection{Cauchy-Characteristic Matching for 3D Scalar Waves}

CCM has been successfully implemented in the fully 3D problem of
nonlinear scalar waves evolving in a flat space-time~\cite{jcp97,Ccprl}.
This study demonstrated the feasibility of
matching between Cartesian Cauchy coordinates and spherical null
coordinates, the setup required to apply CCM to the binary
black hole problem. Unlike the previous examples of matching, the
Cauchy and characteristic patches did not now share a common coordinate
which can be used to define the matching interface. This introduced a
major complication into the matching procedure, resulting in extensive
use of inter-grid interpolation. The accompanying short wavelength
numerical noise presented a new challenge in obtaining a stable
algorithm.

The nonlinear waves were modeled by the equation
\begin{equation}
   c^{-2}\partial_{tt}\Phi = \nabla^{2}\Phi+F(\Phi)+S(x,y,z,t),
   \label{eq:swe}
\end{equation}
with self-coupling $F(\Phi)$ and external source $S$.  The initial
Cauchy data $\Phi(x,y,z,t_0)$ and $\partial_t\Phi(x,y,z,t_0)$ are
assigned in a spatial region bounded by a spherical matching surface of
radius $R_m$.

The characteristic initial value problem (\ref{eq:swe}) is expressed in
standard spherical coordinates $(r,\theta,\varphi)$ and retarded time
$u=t-r+R_m$:
\begin{equation}
    2\partial_{ur}g = \partial_{rr}g - {{L^2 g} \over r^{2}}
       + r(F+S),
\label{eq:SWE}
\end{equation}
where $g = r\Phi$ and $L^2$ is the angular momentum operator
\begin{equation}
    L^2 g = - {{\partial_{\theta}(\sin \theta \partial_{\theta}g)}
    \over {\sin \theta}}
  -{{\partial_{\varphi \varphi}g} \over {\sin^2 \theta}}.
\end{equation}
The initial null data consist of $g(r,\theta,\varphi,u_0)$ on the outgoing
characteristic cone $u_0 =t_0$ emanating from the matching worldtube at the
initial Cauchy time.

CCM was implemented so that, in the continuum limit, $\Phi$ and its normal
derivatives would be continuous across the interface $r= R_{m}$ between the
regions of Cauchy and characteristic evolution. The use of a Cartesian
discretization in the interior and a spherical discretization in the exterior
complicated the treatment of the interface. In particular, the stability of the
matching algorithm required careful attention to the details of the inter-grid
matching. Nevertheless, there was a reasonably broad range of discretization
parameters for which CCM was stable.

Two different ways of handling the spherical coordinates were used. One was
based upon two overlapping stereographic grid patches and the other upon a
multiquadric approximation using a quasi-regular triangulation of the sphere.
Both methods gave similar accuracy. The multiquadric method showed a slightly
larger range of stability. Also, two separate tactics were used to implement
matching, one based upon straightforward interpolations and the other upon
maintaining continuity of derivatives in the outward null direction (a
generalization of the Sommerfeld condition). Both methods were stable for a
reasonable range of grid parameters. The solutions were second order accurate
and the Richardson extrapolation technique could be used to accelerate
convergence.

The performance of CCM was compared to traditional ABC's.  As expected, the
nonlocal ABC's yielded convergent results only in linear problems, and
convergence was not observed for local ABC's, whose restrictive assumptions
were violated in all of the numerical experiments. The computational cost of
CCM was much lower than that of current nonlocal conditions. In strongly
nonlinear problems, matching appears to be the only available method which is
able to produce numerical solutions which converge to the exact solution with a
fixed boundary.
 
\subsection{3D Cauchy-Characteristic Matching}
\label{sec:3dccm}

The most important application of CCM is anticipated to be the binary
black hole problem. The 3D Cauchy codes now being developed to solve
this problem employ a single Cartesian coordinate
patch \cite{Alliance97b}. A thoroughly tested and robust 3D
characteristic code is now in place \cite{high}, ready to match to the
boundaries of this Cauchy patch. Development of a stable
implementation of CCM represents the major step necessary to
provide a global code for the binary problem.

From a cursory view, the application of CCM to this problem might seem
routine, tantamount to translating into finite difference form the
textbook construction of an atlas consisting of overlapping coordinate
patches. In practice, it is an enormous project.

A CCM module has been constructed and interfaced with Cauchy and characteristic
evolution modules. It provides a model of how Cauchy and characteristic codes
can be pieced together as modules to form a single global code. The
documentation of the underlying geometrical algorithm is given in Ref.
\cite{vishu}. The main submodules of the CCM module are:

\begin{itemize}

\item The {\bf outer boundary module} which sets the grid structures.
This defines masks identifying which points in the Cauchy grid are to
be evolved by the Cauchy module and which points are to be interpolated
from the characteristic grid, and vice versa. The reference base for
constructing the mask is the matching worldtube, which in Cartesian
coordinates is the ``Euclidean'' sphere $x^2+y^2+z^2=R^2$. The choice
of lapse and shift for the Cauchy evolution govern the dynamical
and geometrical properties of the matching worldtube.

\item The {\bf extraction module} whose input is Cauchy grid data in
the neighborhood of the worldtube and whose output is the inner
boundary data for the exterior characteristic evolution. This module
numerically implements the transformation from Cartesian ``3+1''
coordinates to spherical null coordinates.  The algorithm makes no
perturbative assumptions and is based upon interpolations of the Cauchy
data to a set of prescribed points on the worldtube. The metric
information is then used to solve for the null geodesics normal to the
slices of the worldtube. This provides the Jacobian for the
transformation to null coordinates in the neighborhood of the
worldtube. The characteristic evolution module is then used to
propagate the data from the worldtube to null infinity, where the
waveform is calculated.

\item The {\bf injection module} which completes the interface by using the
exterior characteristic evolution to supply the outer boundary condition for
Cauchy evolution. This is the inverse of the extraction procedure but must be
implemented outside the worldtube to allow for overlap between Cauchy and
characteristic domains. The overlap region is constructed so that it shrinks to
zero in the continuum limit. As a result, the inverse Jacobian can be obtained
to prescribed accuracy in terms of an affine parameter expansion of the null
geodesics about the worldtube.

\end{itemize}

The CCM module has been calibrated to give a second order accurate
interface between Cauchy and characteristic evolution modules. When
its long term stability has been established, it will provide an
accurate outer boundary condition for an interior Cauchy evolution by
joining it to an exterior characteristic evolution which extracts the
waveform at infinity.

\subsection{The Binary Black Hole Inner Boundary}
\label{sec:bbhib}

It is clear that the 3-dimensional inspiral and coalescence of black
holes challenges the limits of present computational know-how. CCM
offers a new approach for excising an interior trapped region which
might provide the enhanced flexibility to solve this problem. In a
binary system, there are major computational advantages in posing the
Cauchy evolution in a frame which is co-rotating with the orbiting
black holes. Such a description seems necessary in order to keep the
numerical grid from being intrinsically twisted. In this
co-orbiting description, the Cauchy evolution requires an inner
boundary condition inside the black holes and also an outer boundary
condition on a worldtube outside of which the grid rotation is likely
to be superluminal. An outgoing characteristic code can routinely
handle such superluminal gauge flows in the exterior \cite{high}. Thus,
successful implementation of CCM would solve the exterior boundary
problem for this co-orbiting description.

CCM also has the potential to handle the two black holes inside the Cauchy
region. As described earlier with respect to Fig. \ref{fig:1dexci}, an ingoing
characteristic code can evolve a moving black hole with long term stability
\cite{excise,wobb}. This means that CCM might also be able to provide the inner
boundary condition for Cauchy evolution once stable matching has been
accomplished. In this approach, the interior boundary of the Cauchy evolution
is located {\it outside} the apparent horizon and matched to a characteristic
evolution based upon ingoing null cones.  The inner boundary for the
characteristic evolution is a trapped or marginally trapped surface, whose
interior is excised from the evolution.

In addition to restricting the Cauchy evolution to the region outside the black
holes, this strategy offers several other advantages. Although, finding a
marginally trapped surface on the ingoing null hypersurfaces remains an
elliptic problem, there is a natural radial coordinate system $(r,\theta,\phi)$
to facilitate its solution.  Motion of the black hole through the grid reduces
to a 1-dimensional radial problem, leaving the angular grid intact and thus
reducing the computational complexity of excising the inner singular region.
(The angular coordinates can even rotate relative to the Cauchy coordinates in
order to accommodate spinning black holes.) The chief danger in this approach
is that a caustic might be encountered on the ingoing null hypersurface before
entering the trapped region. This is a gauge problem whose solution lies in
choosing the right location and geometry of the surface across which the Cauchy
and characteristic evolutions are matched. There is a great deal of flexibility
here because the characteristic initial data can be posed without constraints.

\begin{figure}[hptb]
  \def\epsfsize#1#2{.5#1}
  \centerline{\epsfbox{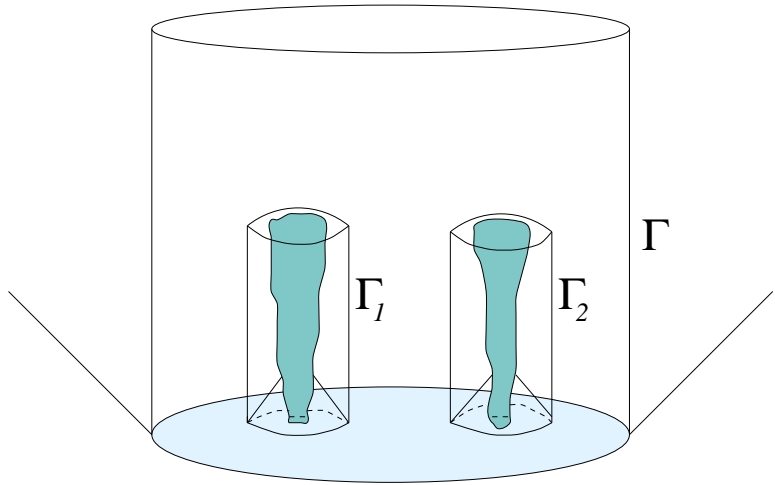}}
  \caption{\it CCM for binary black holes, portrayed in a co-rotating frame.
  The Cauchy evolution is matched across two inner worldtubes $\Gamma_1$ and
  $\Gamma_2$ to two ingoing null evolutions whose inner boundaries excises the
  individual black holes. The outer Cauchy boundary is matched across the
  worldtube $\Gamma$ to an outgoing null evolution extending to ${\cal I}^+$.}
  \label{fig:canbbh}
\end{figure}

This global strategy is tailor-made to treat two black holes in the co-orbiting
gauge, as illustrated in Fig. \ref{fig:canbbh}. Two disjoint characteristic
evolutions based upon ingoing null cones are matched across worldtubes to
a central Cauchy region. The interior boundary of each of these interior
characteristic regions border a trapped surface. At the outer boundary of
the Cauchy region, a matched characteristic evolution based upon outgoing null
hypersurfaces propagates the radiation to infinity.

Present characteristic and Cauchy codes can handle the individual
pieces of this problem. Their unification appears to offer the best
chance for simulating the inspiral and merger of two black holes.
The CCM module is in place and calibrated for accuracy. The one missing
ingredient is its long term stability, which would make future
reviews of this subject very exciting.

\section{Numerical Hydrodynamics on Null Cones}.
\label{sec:grace}

Numerical evolution of relativistic hydrodynamics has been traditionally carried
out on spacelike Cauchy hypersurfaces.  Although the Bondi-Sachs evolution
algorithm can easily be extended to include matter~\cite{isaac}, the advantage
of a light cone approach for treating fluids is not as apparent as for a
massless field whose physical characteristics lie on the light cone. However,
preliminary results from recent studies of a fluid moving in the vicinity of a
black hole indicate that this approach can provide accurate simulations of mass
accretion and the gravitational radiation from an infalling body.

In a 3-dimensional study, a naive hydrodynamic code for a perfect fluid was
incorporated into the PITT null code~\cite{matter}. The fully nonlinear
3-dimensional matter-gravity null code was tested for stability and accuracy to
verify that nothing breaks down as long as the fluid remains well behaved, e.g.
hydrodynamical shocks do not form. The code was used to simulate a localized
blob of matter falling into a black hole, verifying that the motion of the
center of the blob approximates a geodesic and determining the waveform of the
emitted gravitational radiation at ${\cal I}^+$. The results establish the
feasibility of a characteristic {\it matter plus gravity} evolution.

This simulation was a prototype of a neutron star orbiting a black hole. It
would be unrealistic to expect that a naive fluid code would allow evolution of
a compact star for several orbits as it spiraled into a black hole. However, a
refined characteristic hydrodynamic code would open the way to explore this
important astrophysical problem. Recently, J.~A. Font and P.
Papadopoulos~\cite{toniphi} successfully implemented such a code in the case of
spherical symmetry, using the Bondi-Sachs formalism as adapted to describe
gravity coupled to matter inside a worldtube boundary \cite{tam}.

The approach is based upon a high resolution shock-capturing (HRSC) version of
relativistic hydrodynamics developed by the Valencia group (for a  review see
Ref.~\cite{tonirev}). In the HRSC scheme, the hydrodynamic equations are
written in flux conservative, hyperbolic form. In each computational cell, the
system of equations is diagonalized to determine the characteristic fields and
velocities and the local Riemann problem is solved to obtain a solution
consistent with physical discontinuities. This allows a finite differencing
scheme along the fluid characteristics that preserves the physically conserved
quantities and leads to a stable and accurate treatment of shocks. Because the
general relativistic system of hydrodynamical equations is formulated in
covariant form, it can equally well be applied to spacelike or null foliations
of the spacetime. The null formulation gave remarkable performance in the
standard Riemann shock tube test carried out in a Minkowski background and gave
second order convergence in curved space tests based upon
Tolman-Oppenheimer-Volkoff equilibrium solutions. In the dynamic
self-gravitating case, simulations of spherical accretion of a fluid onto a
black hole were stable and free of any numerical problems. Accretion was
successfully carried out in the regime where the mass of the black hole
doubled. In more recent work, they have used the code to study how accretion modulates both decay rates and oscillation frequencies of the 
quasi-normal modes of the interior black hole~\cite{imprints}.

Plans are being made to combine the characteristic evolution codes
for vacuum spacetimes and HRSC relatvitistic
hydrodynamics.

\section{Acknowledgments}
\label{acknow}
This work was partially supported by NSF PHY 9988663. I want to thank the many
people who have supplied me with references. Please keep me updated.

\bibliography{refs}

\end{document}